\PassOptionsToPackage{prologue,dvipsnames,table}{xcolor}
\documentclass[sigconf,divpsnames]{acmart}
\renewcommand\footnotetextcopyrightpermission[1]{}
\AtBeginDocument{%
  }

\setcopyright{acmlicensed}
\copyrightyear{2025}
\acmYear{2025}
\acmDOI{10.5281/zenodo.14396613}
\acmConference[2025]{Conference}{Conference}

\acmISBN{978-1-4503-XXXX-X/18/06}
\usepackage{mdframed}
\usepackage{svg}

\usepackage{xspace}
\usepackage{paralist}

\usepackage{tabularx}
\usepackage{makecell}
\usepackage[acronym,nomain,nonumberlist]{glossaries}

\usepackage[font=small,labelfont=bf,skip=2pt]{caption}
\makeglossaries

\newacronym{ai}{AI}{artificial intelligence}
\newacronym{bsp}{BSP}{bulk synchronous parallel}
\newacronym{mpi}{MPI}{Message Passing Interface}
\newacronym{hpc}{HPC}{high performance computing}
\newacronym{aws}{AWS}{Amazon Web Services}
\newacronym{gke}{GKE}{Google Kubernetes Engine}
\newacronym{aks}{AKS}{Azure Kubernetes Service}
\newacronym{eks}{EKS}{Elastic Kubernetes Service}
\newacronym{ml}{ML}{machine learning}
\newacronym{rdma}{RDMA}{Remote Direct Memory Access}
\newacronym{vm}{VM}{virtual machine}
\newacronym{fom}{FOM}{figure of merit}
\newacronym{efa}{EFA}{Elastic Fabric Adapter}
\newacronym{ec2}{EC2}{Elastic Compute Cloud}
\newacronym{ucx}{UCX}{Unified Communication X}
\newacronym{cni}{CNI}{container networking interface}

 \usepackage[compact]{titlesec}         % you need this package
\newcommand{\low}{\cellcolor{LimeGreen!70} low}
\newcommand{\medium}{\cellcolor{BurntOrange!70} medium}
\newcommand{\high}{\cellcolor{Red!80} high}

\begin{document}

% Idea from Tapasya:
% survey of performance and usability of HPC applications in cloud
\title{Usability Evaluation of Cloud for HPC Applications}
% \title{Cross Cloud Performance Study\protect\\ Informing Possible Futures for HPC\protect\\}

% \subtitle{\large(1) Paper Type: Regular}


\author{Vanessa Sochat}
\authornote{Corresponding Author}
\email{sochat1@llnl.gov}
\orcid{0000-0002-4387-3819}
\affiliation{%
  \institution{\normalsize Lawrence Livermore National Laboratory}
  \city{Livermore}
  \state{California}
  \country{USA}
}

\author{Daniel Milroy}
\email{milroy1@llnl.gov}
\orcid{0000-0001-6500-3227}
\affiliation{%
  \institution{\normalsize Lawrence Livermore National Laboratory}
  \city{Livermore}
  \state{California}
  \country{USA}
}

\author{Abhik Sarkar}
\email{sarkar6@llnl.gov}
\orcid{0009-0007-6733-0548}
\affiliation{%
  \institution{\normalsize Lawrence Livermore National Laboratory}
  \city{Livermore}
  \state{California}
  \country{USA}
}

\author{Aniruddha Marathe}
\email{marathe1@llnl.gov}
\orcid{0000-0003-0546-4472}
\affiliation{%
  \institution{\normalsize Lawrence Livermore National Laboratory}
  \city{Livermore}
  \state{California}
  \country{USA}
}

\author{Tapasya Patki}
 \email{patki1@llnl.gov}
 \orcid{0000-0003-0546-4472}
 \affiliation{%
   \institution{\normalsize Lawrence Livermore National Laboratory}
   \city{\normalsize Livermore}
   \state{\normalsize California}
   \country{\normalsize USA}
}

\renewcommand{\shortauthors}{Sochat et al.}

% NOTE FROM V: 1/3/2025 I am going to comment out deletions so you can see what I removed.
\begin{abstract}
The rise of AI and the economic dominance of cloud computing have created a new nexus of innovation for high performance computing (HPC), which has a long history of driving scientific discovery. In addition to performance needs, scientific workflows increasingly demand capabilities of cloud environments: portability, reproducibility, dynamism, and automation. As converged cloud environments emerge, there is growing need to study their fit for HPC use cases. Here we present a cross-platform usability study that assesses 11 different HPC proxy applications and benchmarks across three clouds (Microsoft Azure, Amazon Web Services, and Google Cloud), six environments, and two compute configurations (CPU and GPU) against on-premises HPC clusters at a major center. We perform scaling tests of applications in all environments up to 28,672 CPUs and 256 GPUs. We present methodology and results to guide future study and provide a foundation to define best practices for running HPC workloads in cloud.
\end{abstract}

\received{20 February 2025}
\received[revised]{12 March 2025}
\received[accepted]{5 June 2025}

\maketitle
\section{Introduction}

% Note from V: I removed "computing" innovation because I think innovation extends beyond just computing. I also changed "deliver products" to just "products" since it's reasonable to carry forward "produce." I also removed "traditional" HPC simulations because I don't see why a simulation needs to have that label.

Due to its rapid growth and economic might, cloud is expected to continue as a primary driver of computing innovation, growing at 20\% per year to reach over \$1.28T USD of revenue by 2028~\cite{gartner-2024}. In comparison, traditional, on-premises \gls*{hpc} had revenue of \$60B USD in 2024, and is projected to grow to \$100B USD by 2028 \cite{hyperion-hpc-ai}. \gls*{hpc} is presented with increasing uncertainty and risks associated with cloud market dominance~\cite{Reed2022-sj}. Engineering costs are now high compared to most computing companies and government budgets~\cite{Reed2022-sj}; they will be paid by the companies with the largest market capitalization for investments with the largest returns. The investments may shape computing in ways that are less suitable to scientific computing, extending beyond market share into the ability to hire talent and obtain resources.

As the traditional, on-premises \gls*{hpc} market shrinks relative to cloud and development proceeds proportionally, scientific workflows may be required to use resources not designed for their needs. A lack of understanding of these new environments could prevent workflow deployment and hinder scientific progress. ``Converged computing’’ is a movement that precludes a future mismatch by promoting the integration of \gls*{hpc} and cloud technologies and cultures into a best-of-both-worlds~\cite{cise-special-issue-converged-computing}.
The movement is bidirectional, bringing components of cloud to on-premises setups~\cite{sochat2024hpcalongsideuserspacekubernetes} and \gls*{hpc} to cloud~\cite{slinky-slurm,Sochat2024-the-flux-operator,Milroy2022-pv}.
% Note that some this is from the original intro I wrote for the converged computing LDRD report, which I figure is OK to reuse since it's not published anywhere.
% In pursuit of higher fidelity modeling of more complex systems, 
Scientific workflows routinely integrate \gls*{hpc} simulations with \gls*{ai} and \gls*{ml}. In turn, \gls*{ai} deployments in hyperscaler clouds have adopted \gls*{hpc} practices to improve model performance and efficiency. Both require continuous innovation to generate discoveries and products.

% I removed HPC-native from the above because we haven't introduced it yet.

% - Define that the simulations are one component of a more complex workflow, the part that is hardest to port is the simulation that was designed for HPC native infrastructure.

% A citation would help here
The increasing ubiquity of cloud and its integration of traditional \gls*{hpc} characteristics into converged environments (as well as the reverse) have % turned 
made the usage of  \hspace{1pt} ``cloud'' and ``\gls*{hpc}'' monikers murky. For the purposes of our work, traditional \gls*{hpc}, or ``\gls*{hpc}-native'' is a bare-metal (i.e., not virtualized) environment that enables the highest performance, lowest latency (network, memory access, etc.), and highest parallel efficiency of tightly coupled applications. \gls*{hpc}-native clusters are typically designed to support the execution of a small number of scientific applications, and the cluster software and hardware are optimized for those applications. \gls*{hpc}-native jobs are executed by submission to resource managers and schedulers such as Slurm~\cite{yoo2003slurm}, LSF~\cite{ibm-lsf}, PBS~\cite{pbs}, and Flux~\cite{ahn-2014,AHN2020202}. 

``Cloud'' or cloud-native environments enable automation, portability, resiliency, reproducibility, elasticity, and resource dynamism. They provide access to infrastructure, hardware, and software as paid services. Applications in cloud typically run on provisioned virtualized, containerized, or orchestrated setups provided by hyperscalers. Cloud-native applications may be ported to \gls*{hpc}-native clusters or HPC-native applications to cloud after substantial effort.
%such as Azure CycleCloud, \gls{aws} ParallelCluster, and Google Cluster Toolkit. Their execution in cloud-native environments is relatively novel.

Modern workloads require much of the spectrum between pure, \gls*{hpc}-native and pure cloud-native environments. For example, a composite scientific workflow can contain a tightly coupled scientific simulation and database along with \gls*{ai} services. While cloud-native deployments of \gls*{ai} models with services are well-established, porting components initially designed for \gls*{hpc}-native infrastructure to cloud is less common. The best-suited cloud environments from a performance \textit{and} usability perspective are poorly defined.
% Note from V, 4/10/2025, I changed "largely unknown" to "poorly defined" because I think we want to say "This work hasn't been done and is important" without making a hugely global statement that there exists no work to look at cloud performance.

% As \gls{hpc}-native environments give way to converged computing and portability across infrastructures becomes increasingly desirable, understanding the performance characteristics and usability of tightly coupled applications in converged and cloud-native environments is correspondingly necessary.

Workflow portability has become a component of performance. Portability enables access to a larger pool of resources, allowing users to decide when, how, and where to run their applications based on resource availability, cost, and characteristics. As portability across infrastructures becomes increasingly desirable, understanding the usability and performance characteristics of tightly coupled applications in converged and cloud-native environments is correspondingly necessary, and is the goal of this work. % The goal of this work is to assess the usability and measure and compare the performance of tightly coupled applications running at a scale of over 100 nodes and over 100 GPUs across converged and cloud-native environments running on hyperscalers and \gls{hpc}-native environments.

\subsection{Background}
%\vspace*{-1mm}
Computing infrastructure and software environments are designed to support customer use cases. The cloud customer base is much of the world population, and its use cases are equally diverse. Services designed to meet those needs must be planet-scale, elastic, resilient, and automated. \gls*{hpc} has advanced application performance and scalability to meet the needs of researchers and engineers, and its infrastructure and environment reflect that specificity.

Cloud computing emerged in 2006 with Amazon's first commercial cloud~\cite{Davies2021-cm}, which offered compute as a service. As the prevalence of cloud computing increased, \gls*{hpc} users began testing and running workloads in the cloud, an important development for those without access to centers \cite{access-initiative}. Initial performance studies revealed limitations, particularly with respect to network performance and I/O~\cite{magellan-2011}. 

As cloud services became integral to everyday life, application and infrastructure automation, autoscaling, and resilience assumed increasing importance. Kubernetes~\cite{kubernetes} surfaced to address these challenges as an open source evolution of Google Borg. Kubernetes has become the standard full-lifecycle orchestrator for containerized applications and benefits from over 94K code contributors~\cite{kubernetes-contributors-4-10-2025}. Kubernetes' massive contributor base and ubiquity has made it a lingua franca and portability layer for cloud-native applications. Large-scale, complex and dynamic scientific workflows that integrate traditional \gls*{hpc} simulation together with cloud-oriented components such as \gls*{ai}/\gls*{ml}, messaging, and database services \cite{mummi-2019,mummi-2021,ams-2023,peterson-2022,ahn-2022,lee-2019,AdditiveFOAM_1.0.0,montage-2014,siva_1000_2008,cybershake-2024} can benefit from this ecosystem.
\gls*{hpc} applications running in containerized environments have not been observed to experience performance degradation compared to bare-metal execution~\cite{younge-2017,liu-2022,Hu2019-ip}, and \gls*{rdma} and OS-bypass mechanisms avoid overhead introduced by the Kubernetes network \cite{Beltre2019-mr}.
% bypass the Kubernetes network that can introduce overhead \cite{Beltre2019-mr}. 
With these mechanisms in place, the environment overhead is potentially small. 

Understanding the practical usability, benefits, relative overheads, and performance limitations of scientific applications running in these environments can direct more traditional performance studies, and provide real-world experience to better inform workflow movement between infrastructures.

\vspace{3pt}
\noindent{\bf Related Work. } 
Recent studies test \gls*{mpi}-based applications in Kubernetes~\cite{Milroy2022-pv, Sochat2024-the-flux-operator} or in a single or small number of environments \cite{Gupta2011-uq, Chang2018-jd}. Performance and usability studies in the last decade have equivalently looked at small numbers of nodes and applications \cite{Gienger2015-zc}. Recent studies have tested applications at larger scales on a limited number of environments ~\cite{lange-2023,dancheva2023,Milroy2022-pv}. Numerous studies investigate performance and usability and describe the porting process of small numbers of \gls*{hpc} applications running in the cloud at smaller scales. Surveys like~\cite{netto2018} provide an overview of such works. We are not aware of studies that evaluate usability and application performance of over 10 applications and on multiple environments, including on-premises, \gls*{vm}, and Kubernetes, at a scale of over 100-nodes and 100-GPUs.

\vspace{3pt}
\noindent{\bf Contributions. } 
For this study, we aimed to build a suite of 11 CPU and GPU \gls*{hpc} applications to study performance and usability of 14 environments. We targeted the most commonly used clouds and viable setups offered by each. We used qualitative and quantitative metrics, evaluating the difficulty of setup and execution, and accounting for extra development or debugging needed.  We tested features (e.g., auto-scaling) that are not native to HPC, but suggested for specific cloud environments.  Our contributions include:

\begin{compactitem}
 \item{Suggested practices for orchestration of cloud studies}
 \item{Qualitative usability assessment of 12 cloud environments}
  \item{Synthetic benchmark and proxy app performance on up to 256 nodes and GPUs across 13 different environments}
 \item{Kubernetes configurations of 11 HPC apps on 32-256 nodes}
 \item{Daemonset to deploy Infiniband to Azure}
 \item{97 containerized CPU/GPU HPC apps specific to each cloud}
 \item{Azure \gls{vm} bases for Ubuntu 24.04 with NVIDIA drivers}
 \item{25,541 datasets (runs), 3,546 reported in the paper.}
\end{compactitem}

Performing a more comprehensive evaluation of cloud environments is a challenging task that is notably different from running studies on HPC; along with needing custom setups and builds, cloud environments are ephemeral, often opaque to the user, and change over time. Due to resource contention and budget limits, testing is often limited or not possible. However, such an understanding can prompt and facilitate collaboration between \gls*{hpc} users, researchers and cloud engineers. 

We first provide an overview of our methods (Section \ref{sec:methods}), discussing the practical steps to obtain resources, software, applications, and environment setups. In our results (Section \ref{sec:results}) we investigate environment usability and application performance. We conclude with a discussion of usability, suggested practices, and feedback for each cloud in anticipation of future work (Section \ref{sec:discussion}). We contribute to the HPC community a practical understanding of environments available, and suitability for different scales of work.

% Counts for experiments (these are what go into final figures, after filter)
% AMG total number of CPU datum:    201
% AMG total number of GPU datum:    226
% KRIPKE total number of CPU datum: 146
% LAGHOS total number of CPU datum:  28
% LAMMPS total number of CPU datum: 132
% LAMMPS total number of GPU datum: 118
% minife total number of CPU datum:  82
% minife total number of GPU datum: 185
% mixbench: 9465 result files across experiments, BUT we only report ECC, which we get from 384 GPU node results.
% mt-gemm Total number of CPU datum: 100 NOT USED
% mt-gemm Total number of GPU datum: 114

% OSU total number of CPU datum:     991
% Quicksilver total number of CPU datum: 129
% Stream has 10,367 file results... lol BUT we only use 810 in our paper.
% I'm not going to count the single node benchmark (3259 runs)
% TOTAL (for now): 3546
% If we count all data (mixbench, single node, and stream): 25541

\section{Methods}
\label{sec:methods}

\subsection{Overview of Study Design}
\label{sec:study-design}

The study aimed at a best-effort assessment of usability and performance of different cloud vendor offerings for environments that support \gls*{hpc} workloads, including those provided by \gls*{aws}, Google Cloud, and Microsoft Azure at the largest scale possible given our budget. Usability in this context is primarily qualitative and includes an assessment of ease of obtaining and using resources and supporting tools, along with the extent to which additional development or debugging is needed. Performance is measured by the reported or derived \gls*{fom} specific to each application. We budgeted for a spend of \$49,000 USD per cloud and targeted a scale of 256 CPU nodes and 256 GPUs. Due to the high rate of change of cloud resources and technologies, the data collected is a snapshot of clouds' states with a shorter term of representation than is typical for \gls*{hpc} studies. Throughout this paper we use the terms ``nodes'' and ``instances'' interchangeably.

\subsection{Architecture and Hardware}
\label{section:architecture}

Comparable architectures and hardware were needed to make a fair comparison across clouds and local environments. Our goal was to use the newest generation of GPUs with an available quantity of 256 GPUs (8 GPU/node corresponds to a 32 node cluster). We chose the GPU count to represent a moderate-scale \gls*{hpc} application execution and to provide a large enough maximum size to test the application scaling characteristics. We adopted a similar approach to testing CPU nodes, choosing instances that offered a comparable architecture based on \emph{AMD EPYC}\texttrademark ~CPUs available in quantities of 256 or more. Within this set of comparable CPU architectures, AWS and Azure offered 96 cores per instance, while Google Cloud had 56, a difference that should be noted when considering results.

It was not possible to obtain newer generations of GPUs in our desired quantity on any cloud. % Obtaining more than a few nodes with \emph{H100}s or \emph{A100}s on Amazon would not be possible. On Google Cloud \emph{A100}s were reported to be severely constrained, and \emph{H100}s were just coming online. Microsoft could offer an estimated 5 nodes of \emph{A100}, also due to high demand. In light of these constraints, 
The three-generation-old \emph{NVIDIA V100} was the only way to do a comparison with the same hardware across clouds at our desired scale. While using an older model is not ideal, it would provide a fair comparison of equivalent hardware, and it reduced risk, as many of our study applications were well-tested on \emph{NVIDIA V100}s but not on newer GPUs. 
% After creation of our cloud accounts and discussing resource availability for each of CPU and GPU, we requested resource quota \cite{google-cloud-quotas-and-limits}, a semi-manual process that starts with an interface requests and typically is completed over call or email. 
Complete details for all nodes used in the study are included in Table~\ref{table:node-characteristics}. % The reality that we faced -- not being 

\subsection{Experiment Environments}

\smallskip
\noindent{\bf Kubernetes. } 
We tested Kubernetes, the de facto standard cloud orchestrator~\cite{kubernetes} using the \gls*{eks} on Amazon (v1.27), \gls*{aks} on Microsoft (v1.29.7), and \gls*{gke} on Google Cloud (v1.29.7). 
%\gls*{gke} uses the Container Optimized OS \cite{google-cos} with Kubernetes version 1.29.7, \gls*{eks} uses the ``amazon/amazon-eks-node-1.27'' (Kubernetes 1.27) image with Amazon Linux 2, and \gls*{aks} uses the ``AKSUbuntu-2204gen2'' image with Kubernetes 1.29.7. 
% Kubernetes has desirable features including resource dynamism, elasticity, resiliency, portability, and automated management that can greatly benefit emerging, complex HPC workloads. 
We further unified our experiments by using the \emph{Flux Operator} \cite{Sochat2024-the-flux-operator}, a Kubernetes operator that deploys a Flux cluster % called a ``MiniCluster'' 
across nodes. While Flux was initially developed for on-premises \gls*{hpc}, it is agnostic to the nodes used, and runs on cloud \gls*{vm}s \cite{AHN2020202}. % This is an example of convergence, as it provides a translation layer between HPC and cloud. A workload that runs on-premises with Flux can be easily ported to a cloud environment also running Flux. 

\medskip
\noindent{\bf Virtual Machines. } 
Each cloud advertises different \gls*{vm} solutions that are suitable for \gls*{hpc} workloads, including ParallelCluster (\gls*{aws}) \cite{yoo2003slurm}, Cluster Toolkit (Google Cloud), and CycleCloud (Microsoft Azure). These setups extend the capability of traditional \gls*{hpc} clusters by offering the ability to add and remove nodes dynamically. In these environments, we opted to maximize the comparability of the experiments with Kubernetes by using the same containers, but pulled directly to the machines using Singularity~\cite{Kurtzer2017-xj}. % Containers not only reduce development work -- they have been shown many times to be equal to if not better than bare-metal performance \cite{polgar2023three,keller2023containers}.  
% which then would be available to spawned worker nodes.  For Google Cloud, we tested setups using Slurm and Flux provided by Cluster Toolkit, and ran into errors installing GPU drivers. We used a derivative \cite{vanessa_2024_14396613} of the Flux recipe provided in the Google Scientific Examples repository \cite{google-scientific-computing-examples} that also is the basis for the Cluster Toolkit version. This setup brings up a cluster of one size, without autoscaling, and so containers were pulled to (and built directly into) the virtual machine image deployed as the nodes.

% Note to others: it mostly only varies based on the VM build (I reduced 15+ files to a single build script
% see original https://github.com/GoogleCloudPlatform/scientific-computing-examples/tree/main/fluxfw-gcp/img
% and then building just one image to use across nodes (Ward built a manager, login, and worker). Note that Ward's setup does not max out MTU (1460 instead of 8896) but I tested the setup with high MTU later and it didn't meaningfully alter the result.

\medskip
\noindent{\bf On-premises. } 
We used the % Dane and Lassen 
\emph{A} and \emph{B} clusters at \emph{Institution} for CPU and GPU runs, respectively. % \cite{llnl-dane,llnl-lassen}. 
\emph{A} (2023) has 1,544 total nodes with \emph{Intel Xeon} CPUs and 112 cores per node, and provides the Slurm workload manager \cite{jette2023architecture}. \emph{B} is an IBM system (2018) that provides 795 total nodes with 44 \emph{IBM POWER9} CPUs and 4 \emph{NVIDIA V100} GPUs per node. It uses LSF \cite{ibm-lsf} for job submission and scheduling. % More on the technical aspects of these systems are discussed in Section \ref{section:architecture}.

\subsection{Scale}
\label{section:scale}

For CPU experiments we tested cluster sizes of 32, 64, 128, and 256 nodes. For GPU experiments we sought to provision cluster sizes of 4, 8, 16, and 32 nodes, each with 8 NVIDIA V100 GPUs. Due to the different number of GPU per node on \emph{B} (4) as compared to cloud (8) we compare sizes 8, 16, 32, and 64 nodes on \emph{B} to sizes 4, 8, 16, and 32 on cloud, recognizing that \emph{B} would require greater network utilization given twice the number of nodes at each size. % It also differed in using IBM Spectrum LSF \cite{quintero2015ibm} as a workload manager.  
% For CPU, the most comparable choice for on-premises turned out to be a cluster with Intel Xeon, the <A> cluster. 
Details about nodes (Table~\ref{table:node-characteristics}) and environments (Table~\ref{table:environment-characteristics}) are provided. 

\subsection{Usability Evaluation}

The goal of our usability study was to provide researchers and \gls*{hpc} developers with an understanding of available environments, and to provide clouds with feedback to address shortcomings and improve user experience. For each environment we aimed to qualitatively evaluate the sequence of steps between setup and execution. Specific steps and tasks for each cloud environment are available \cite{vanessa_2024_14396613}, and include setup (testing, deployment, and configuration), development (addressing needs identified for setup), application setup (building containers, environment, and parameters), and manual intervention (interactions and monitoring required to orchestrate the study). For each category and environment, we will provide a qualitative measure of effort as was done in \cite{Gienger2015-zc} that reflects our subjective experience of the step. A value of ``low'' indicates that procedure required for the step worked after following available instructions, with minimal development or configuration needed. A value of ``medium'' reflects facing unexpected issues needing debugging or development, and a value of ``high'' means that there was significant development effort. We also aimed to provide an assessment of the difficulty to obtain accounts and resource quotas.

%\begin{figure}
%  \includesvg[width=300pt]{images/usability-table.svg}
%  \caption{Primary tasks involved for experiment steps. Unless noted, each step is relevant for CPU and GPU.}
%  \label{fig:environment-table}
%\end{figure}

\subsection{Networking}

\gls*{aws} setups used the \gls*{efa} suggested for \gls*{hpc} workloads \cite{amazon-efa} that can be installed in \gls*{ec2} and \gls*{eks} environments to allow for network packets to bypass the operating system and go directly to the device. For ParallelCluster, \gls*{efa} was added via a parameter in the cluster configuration file \cite{parallel-cluster-cpu-efa}. For \gls*{eks}, it was installed via a daemonset.  For Microsoft Azure, all setups used \gls*{hpc} \gls*{vm}s that came with InfiniBand enabled \cite{azure-infiniband}. %, another strategy to enable \gls*{rdma} for \gls*{hpc} workloads. 
There were no comprehensive instructions for enabling the same drivers on \gls*{aks}, requiring development work by our team (Section \ref{sec:usability-development}).
% Note that our initial study did not equate Compute Engine with GKE because of differences in maximum transmission unit (MTU) \cite{mtu} and network Tier. For MTU, our Compute Engine setup used a default of 1460, while we knew how to set and use a higher value of 8896 for GKE. 
\gls*{gke} experiments used Premium Tier 1 networking to improve bandwidth, and Compute Engine clusters used Standard \cite{google-network-tiers}. % Note from V: I did follow up work and found that bumping MTU and using Tier 1 bandwidth had little to no impact to the initial results. There was still a huge difference.

Each cloud allows requesting nodes to be placed in close proximity, typically in the same rack, region, or zone. At the time of our study, Google Cloud \emph{COMPACT} could be requested for up to 150 nodes (recently increased to 1500) to place nodes in the same zone~\cite{google-compact}. On \gls*{aws} a \emph{cluster placement group} ensures that nodes are packed closely together in a single Availability Zone \cite{Amazon_Web_Services2022-placement}. Azure defines \emph{proximity placement groups} \cite{azure-placement} for instances to be created in a single datacenter. This strategy is suited to more tightly coupled workloads. 
Finally, our on-premises clusters used low-latency fabrics \emph{Omni-Path 100} \cite{birrittella2016enabling} on \emph{A} and \emph{InfiniBand EDR} \cite{filliater2012infiniband} on \emph{B}. 

\begin{table}
\small
  \caption{Environment Characteristics}
  \label{table:environment-characteristics}
  \begin{tabularx}{0.85\columnwidth}{lll}%{@{}p{2.9cm}lll@{}}
      \toprule
    Environment & Scheduler & Containers \\
    \midrule
     CPU \emph{A} \textsubscript{(p)} & Slurm & No \\
     CPU ParallelCluster \textsubscript{(aws)} & Slurm & Yes \textsubscript{(s)} \\
     CPU EKS \textsubscript{(k,aws)} & Flux & Yes \textsubscript{(cd)} \\
     CPU Compute Engine \textsubscript{(g)} & Flux & Yes \textsubscript{(s)} \\
     CPU GKE \textsubscript{(k,g)} & Flux & Yes \textsubscript{(cd)}  \\
     CPU CycleCloud \textsubscript{(az)} & Slurm & Yes \textsubscript{(s)} \\
     CPU AKS \textsubscript{(k,az)} & Flux & Yes \textsubscript{(cd)}  \\
     \hline
     GPU \emph{B} \textsubscript{(p)} & LSF & No \\
     GPU ParallelCluster \textsubscript{(aws)} & Slurm & Yes \textsubscript{(s)} \\
     GPU EKS \textsubscript{(k,aws)} & Flux & Yes \textsubscript{(cd)} \\
     GPU Compute Engine \textsubscript{(g)} & Flux & Yes \textsubscript{(s)} \\
     GPU GKE \textsubscript{(k,g)} & Flux & Yes \textsubscript{(cd)} \\
     GPU Cyclecloud \textsubscript{(az)} & Slurm & Yes \textsubscript{(s)}  \\
     GPU AKS\textsubscript{(k,az)} & Flux & Yes \textsubscript{(cd)} \\
  \bottomrule
\multicolumn{3}{l}{\footnotesize Amazon Web Services \textit{aws}, Microsoft Azure \textit{az}, Google Cloud \textit{g}, On-Premises \textit{p}} \\
\multicolumn{3}{l}{\footnotesize Kubernetes \textit{k}, containerd \textit{cd}, Singularity \textit{s}}  \\
\vspace{-3mm}
\end{tabularx}
\end{table}

% Note from V: The GPU frequency didn't have units - I'm assuming also GHz
% Dane is 112 cores, 224 threads,  but it's the only one that had cores
% Also for Dane I changed Intel(R) Xeon(R) to Intel Xeon is there a difference?
% Frequency is max, I assume it can be implied?
\begin{table*}
 \small
  \caption{Nodes and Network}
  \label{table:node-characteristics}
  \begin{tabular}{@{}p{3.4cm}lllllll@{}}
      \toprule
    Environment & Node Type & Processor/GPU & Cores/Frequency & Memory & Network & Cost/Hr \\
    \midrule
     CPU \emph{A} \textsubscript{(p)} & Dell & Intel Xeon Platinum 8480+ & 112/3.8 GHz & 256GB & Omni-Path 100 & -- \\
     CPU ParallelCluster \textsubscript{(aws)} & Hpc6a & AMD EPYC 7R13/7003 & 96/3.6 GHz & 384GB & EFA Gen1.5 & \$2.88 \\
     CPU EKS \textsubscript{(k,aws)} & Hpc6a & AMD EPYC 7R13/7003 & 96/3.6 GHz & 384GB & EFA Gen1.5 & \$2.88 \\
     CPU Compute Engine \textsubscript{(g)} & c2d-standard-112 & AMD EPYC 7B13 & 56/2.45-3.5 GHz & 448GB & Google Premium & \$5.06 \\
     CPU GKE \textsubscript{(k,g)} & c2d-standard-112 & AMD EPYC 7B13 & 56/2.45-3.5 GHz & 448GB & Google Premium, Tier\_1 & \$5.06 \\
     CPU CycleCloud \textsubscript{(az)} & HB96rs v3 & AMD EPYC 7003 & 96/1.9-3.5 GHz & 448GB & InfiniBand HDR & \$3.60 \\
     CPU \gls*{aks} \textsubscript{(k,az)} & HB96rs v3 & AMD EPYC 7003 & 96/1.9-3.5 GHz & 448GB & InfiniBand HDR & \$3.60 \\
     \hline
     GPU \emph{B} \textsubscript{(p)}& IBM & IBM Power9/V100 16GB & 44/3.5 GHz max & 256GB & InfiniBand EDR & -- \\
     GPU ParallelCluster \textsubscript{(aws)} & p3dn.24xlarge & Xeon Platinum 8175/V100 32GB & 48/2.5 GHz & 768GB & EFA Gen1 & \$34.33 \\
     GPU \gls*{eks} \textsubscript{(k,aws)} & p3dn.24xlarge & Xeon Platinum 8175/V100 32GB & 48/2.5 GHz & 768GB & EFA Gen1 & \$34.33 \\
     GPU Compute Engine \textsubscript{(g)} & n1-standard-32 & Xeon Haswell E5 v3/V100 16GB &  16/2.3 GHz & 120GB & Google Premium & \$23.36 \\
     GPU \gls*{gke} \textsubscript{(k,g)} & n1-standard-32 & Xeon Haswell E5 v3/V100 16GB & 16/2.3 GHz & 128GB & Google Premium & \$23.36 \\
     GPU Cyclecloud \textsubscript{(az)} & ND40rs v2 & Xeon Platinum 8168/V100 32GB & 48/2.7 GHz & 672GB & InfiniBand EDR & \$22.03 \\
     GPU \gls*{aks} \textsubscript{(k,az)} & ND40rs v2 & Xeon Platinum 8168/V100 32GB & 48/2.7 GHz & 672GB & InfiniBand EDR & \$22.03 \\
  \bottomrule
\multicolumn{7}{l}{\footnotesize Amazon Web Services \textit{aws}, Microsoft Azure \textit{az}, Google Cloud \textit{g}, On-Premises \textit{p}, Kubernetes \textit{k}. Cost is hourly in USD for a single instance, with GPUs included.} \\
\end{tabular}
\end{table*}

\subsection{Software Builds}

\medskip
\noindent{\bf Containers. } 
All cloud experiments used common containers and base images, with an emphasis on maximizing similarity within and across environments. Containers run in \gls*{vm} environments via Singularity were equivalent to those deployed in Kubernetes. % We used common base containers, with Azure, Google Cloud and \gls*{aws} CPU using ubuntu:22.04, Google Cloud and \gls*{aws} GPU using nvidia/cuda:12.4.1-cudnn-devel-ubuntu22.04, and Compute Engine CPU using rockylinux:9. 
For each container we installed the same versions of Flux Framework projects (flux-security 0.11.0, flux-core 0.61.2, flux-sched 0.33.1, flux-pmix 0.4.0) and core dependencies (cmake 3.23.1, libfabric 1.21.1 for AWS), along with OpenMPI 4.1.2. We chose OpenMPI because it is suggested for use on Azure \cite{openmpi-azure} and \gls*{aws} \cite{aws-openmpi}, and in testing we found it outperformed Intel MPI in Google Cloud \cite{hpckm-sochat}.  We also built the same releases or commits of applications \cite{vanessa_2024_14396613}.  % In the case of a commit, the exact identifier can be found in the ``docker'' directory of the performance study repository \cite{vanessa_2024_14396613} that includes tables of containers, Dockerfiles, and a complete software catalog. 
Containers are deployed to the registry alongside the repository \cite{vanessa_2024_14396613}.%, and the containers associated directly with the same code repository for easy discoverability \cite{vanessa_2024_14396613}.

Differences in container builds came down to drivers and networking software. \gls*{aws} needed OpenMPI compiled with libfabric for \gls*{efa}. Azure containers required the \gls*{ucx} for InfiniBand \cite{Shamis2015-za}. Google Cloud did not need any special software or drivers, and could share containers with \gls*{aws}. Per suggested practice \cite{google-scientific-computing-examples,rocky-linux-optimized-for-google-cloud}, we built Rocky bases for Compute Engine.

\medskip
\noindent{\bf Virtual Machine Images. } 
We used the recommended Rocky Linux Optimized for Google Cloud \cite{rocky-linux-optimized-for-google-cloud} for the Compute Engine base image, installing the same versions for each of the Flux packages, OpenMPI, and Singularity. This \gls*{vm} base used the same set of build instructions as was used for the containers, ensuring that the environments and paths were equivalent. \gls*{aws} ParallelCluster and Azure CycleCloud \gls*{vm}s were provided to us.

\medskip
\noindent{\bf On-premises Builds. } 
Bare-metal builds for \emph{B} (GPU) and \emph{A} (CPU) required building applications directly on the systems using software modules.  CPU and GPU variants of AMG2023 (Section \ref{section:amg}) were built using the Spack package manager \cite{gamblin2015spack,vanessa_2024_14396613}, and all other applications were built from respective open source repositories. % using system level MPI. 
% Environments for Spack are provided \cite{vanessa_2024_14396613}.

\subsection{Applications}
\label{sec:applications}

The choice of our 11 applications was dictated by the CORAL-2 benchmarks~\cite{coral-2-benchmarks} used to assess enterprise systems at DOE centers as well as other common \gls*{hpc} proxy apps and synthetic benchmarks. % We chose a subset (11) of our 12 to present here, as 1 was eliminated due a hard requirement for a shared filesystem. 
For each, strong or weak scaling was chosen based on properties of the application \cite{strong-weak-scaling}. Five iterations were run at each scale. % Strong scaling is governed by Amdahl’s law, and is done by adding resources while a problem size is held constant. An equivalent speedup should be observed, as more resources are available for the same problem size. Weak scaling is governed by Gustafson’s law, and is concerned with changing the problem size as resources are adjusted to achieve the same performance. 
% We reported \gls*{fom}s appropriate for each application.

% For Laghos there were definitely a few runs I went up to 25 minutes just to see if it would end...
% Note that I'm putting these in ABC order for now.

\label{section:amg}
% From docker build notes:  cpu gets +int64, and gpu +mixedint
% Note that int64 in spack corresponds with --enable-bigint
% https://github.com/spack/spack/blob/develop/var/spack/repos/builtin/packages/hypre/package.py
% Dan can you check this over / add details? I did my best but you investigated thoroughly and I probably missed something.
\smallskip
\noindent{\bf AMG2023} required different builds to avoid segfaults. The software uses hypre \cite{hypre}, which has flags for 64-bit and mixed integers. Our GPU builds required the mixed integer flag to set \texttt{HYPRE\_BigInt} to \texttt{long long int}, reducing memory usage and computational costs in comparison with enabling \texttt{long long int} uniformly \cite{hypre, amg2023-github}. The big integer flag was needed for CPU builds to set both \texttt{HYPRE\_BigInt} and \texttt{HYPRE\_Int} to \texttt{long long int} to avoid segfaults and solve larger systems ~\cite{hypre}. For CPU and GPU configurations, we ran problem 2 and used a problem size of 256x256x128 \cite{amg2023-github}. The \gls*{fom} is defined as the following, where nnz\_AP is the total number of non-zeros, and two wall clock times for each of setup and the solving phase \cite{amg2023-github}: $FOM = \frac{nnz\_AP}{Setup Phase Time + 3*Solve Phase Time}$.

We chose a per-GPU problem size that would fit into 16GB GPU memory to be compatible with the \emph{NVIDIA V100} variant offered by Google Cloud and cluster \emph{B}. Our choice also ensured the global problem size was small enough to be indexed by an integer ~\cite{amg2023-github-issue13}. We ran AMG2023 tests in a weak scaling configuration.

\label{section:laghos}
\smallskip
\noindent{\bf Laghos} 
(LAGrangian High-Order Solver) is a proxy app that simulates the compression of gas using a moving Lagrangian frame~\cite{Dobrev2012-mm}. The suggested \gls*{fom} is the major kernels total rate (megadofs $\times$ time steps/second). We ran Laghos in a strong scaling configuration similar to the Vulcan example~\cite{laghos-github} with a \emph{cube\_311\_hex} mesh using partial assembly and a maximum of 400 steps. 

\label{section:lammps-reax}
\smallskip
\noindent{\bf LAMMPS} 
 (Large-scale Atomic/Molecular Massively Parallel Simulator) is a proxy application to model the reaction of atoms, first calculating the force, energy, and pressure during chemical reactions, and then using these calculations to solve a matrix optimization problem. We used the \emph{ReaxFF} package for our study, which refers to a reactive force field~\cite{lammps-reax}. We followed a suggested practice \cite{lammps-reax} to calculate the millions of atom steps per second (m-atom steps) for our \gls*{fom}. A larger value is better, indicating a system can do more calculations per second. We chose a consistent problem size of 64x32x32 to run on GPU, and 64x64x32 for CPU for each of parameters x, y, and z, respectively. The GPU problem size was chosen to be smaller to fit on the GPUs on Google Cloud and \emph{B}. We ran LAMMPS in a strong scaling configuration.

\label{section:kripke}
\smallskip
\noindent{\bf Kripke} 
is a deterministic particle transport proxy application intended for testing how data, programming, and systems impact particle transport performance. It primarily takes parameters that describe energy groups, data layout, and thread and task parallelism and solver setup. We chose to use grind time, the time to complete a unit of work (lower is better), as our \gls*{fom} \cite{kripke}. % While all FOMs might offer interesting insights, for this work, we chose grind time as a metric of interest.

\label{section:minife}
\medskip
\noindent{\bf MiniFE} 
(mini Finite Element proxy application) is a ``proxy application for unstructured implicit finite element codes,'' and is similar to HPCG~\cite{miniFE}. The suggested \gls*{fom} is the total conjugate gradient mega floating-point operations per second (Total CG Mflops) \cite{miniFE}.

\label{section:mt-gemm}
\medskip
\noindent{\bf MT-GEMM}
is part of the NERSC proxy suite (MT-xGEMM) to measure matrix multiplication for GPU \cite{mt-gemm-gpu}. We used an MPI implementation for CPU \cite{mt-gemm-cpu}, part of the PRACE HPC kernels.

\label{section:mixbench}
\medskip
\noindent{\bf Mixbench} 
is a a well-known CPU and GPU benchmark, which we were primarily interested in running as a single-node benchmark to provide basic metrics and attributes for our GPU \cite{mixbench}. % that can provide additional insights.

\label{section:osu}
\medskip
\noindent{\bf OSU Benchmarks} (Ohio State University Benchmarks) are considered an \gls*{hpc} community standard for measuring \gls*{mpi} communication patterns \cite{Paniraja_Guptha2023-rr}. We chose to run point-to-point benchmarks for each of latency (\emph{osu\_latency}) and bandwidth (\emph{osu\_bw}) and the all reduce collective communication pattern (\emph{osu\_allreduce}). For GPU runs, the benchmarks were run using host to host mode (\texttt{cuda -d H H}) as only Infiniband fabrics support GPU Direct (device to device \gls*{rdma}) \cite{shainer2011development}. For point-to-point benchmarks, we opted for a sampling strategy where we randomly selected 8 nodes, and selected at most 28 combinations of pairs to test.

% https://github.com/converged-computing/performance-study/blob/main/docker/google/cpu/single-node/entrypoint.sh
\label{section:single-node}
\medskip
\noindent{\bf Single Node Benchmark} We developed a single node benchmark that would be run on each node and save the output of dmidecode \cite{dmidecode}, CPU information from \texttt{/proc/cpuinfo}, node topology (images and XML)  from hwloc \cite{broquedis2010hwloc}, and output from sysbench \cite{kopytov2004sysbench}. % We built two variants of this container -- one expecting a shared filesystem (for VM environments) that would write to and clean up a temporary space, and a second that did not require it (all Kubernetes environments).

\label{section:stream}
\medskip
\noindent{\bf Stream} is a memory bandwidth benchmark that measures kernels for Copy, Scale, Add, and Triad operations \cite{mccalpin1995stream}. For our study, we use variants for each of CPU \cite{HammondUnknown-ni} and GPU \cite{cuda-stream}, and ran the benchmark in two configurations: single-node (CPU), and across nodes (GPU). To measure memory bandwidth we used the Stream Triad Kernel. %It first multiplies an array by a scalar, followed by an addition to another array, and storing the result in a third array.
This vector kernel operation is common in \gls*{hpc} applications, and is a useful synthetic benchmark for testing memory bandwidth by performing loads, multiplications, and stores. % To test memory bandwidth for GPU, we will run the benchmark on all cluster nodes.

% We supplemented our measurement of Stream on GPU with Mixbench \cite{mixbench}, a well-known GPU benchmark. % that can provide additional insights.
% \subsection{Orchestration}
% Orchestration includes the means to submit jobs (e.g., the workload manager) and save output. % There were two dimensions for the first -- the degree to which a setup supported elasticity, and the workload manager that we used.

\label{section:quicksilver}
\medskip
\noindent{\bf Quicksilver} 
is a proxy application that solves a simplified Monte Carlo particle transport problem~\cite{quicksilver}. It uses OpenMP and MPI and replicates comparable patterns of memory access, communication patterns, and branching and divergence as its full application counterpart~\cite{quicksilver}.

\subsection{Workload Orchestration}

\gls*{eks} was deployed via \emph{eksctl}, \gls*{gke} via \emph{gcloud}, and \emph{aks} via the web interface.
Workloads run in \gls*{vm} environments were deployed following vendor instructions and submit to respective jobs managers. On-premises workloads needed to wait in the queue. The \emph{Flux Operator} \cite{Sochat2024-the-flux-operator} was used in all Kubernetes environments, and each cluster size was deployed independently to be more cost effective.  % we could shell into the lead broker pod to reveal an entire cluster and command line akin to an on-premises cluster. 
% This setup did not require any waiting time for a queue or autoscaling. % was an improvement on both of the above as nodes did not need to be waited for -- we owned the entire set of nodes as the single ``flux'' user, and although the Flux Operator supports elasticity, there was no autoscaling needed. 
% Choosing to run each scale as its own cluster was a decision that we made to be more cost effective, as we would not spend funds on nodes not being used or waiting for resources.  % Although the Flux Operator supports elasticity and autoscaling, we opted for simplicity, and a preference to not need to wait for nodes to come up from the cloud when they were needed.
% While the study did not intentionally include any I/O intensive workloads, we recognize this is an important variable that many workloads, especially those with large AI/ML models, will need to consider. Rather, our primary concern across environments was having strategies to capture and save output, events, and other workload manager metadata.
% Traditional workload managers like Slurm allow specifying that output and error are directed to specific files \cite{slurm-sbatch}. 
% To save output, we echoed timestamps at the onset and end of runs to capture total wrapped runtime for Slurm \cite{slurm-sbatch}. % Given a command to capture a timestamp at the start and the end of a run, this gave us a simple means to capture the total wrapped runtime with the job output, and maintain our own organization of output files. 
Job output was saved to file and pushed to a registry \cite{oras}.
\section{Results}
\label{sec:results}

% This seemed out of place. (1/17/2025)
% We want to note that the cost per hour for the AWS HPC family of instances is currently the most affordable across clouds, with the cheapest (hpc7g, ARM Graviton) being under \$2.00 an hour. 

% \subsection{On-premises Environments}

% Since the on-premises resources were not constrained to the same one month time period of our cloud resources (set by the funding bodies), we opted to do the on-premises runs the following months, in September and October. After our LDRD project ended and jobs were still in the queue, the lower priority meant waiting additional months to complete the runs.  This highlights a challenge with resource availability on a busy, shared HPC multi-tenancy system.
% However, while we were not constrained by funding, we were limited to run in the context of a multi-tenancy queue. When the allocation for our LDRD project expired at the end of October and there were jobs remaining that had not yet run, they took on a very low priority that would deem the work not finished until the next calendar year. 
%  It often is very slow to run workloads having to wait for jobs to reach the top of the queue. If a job is submit erroneously, then the wait has to happen again.

% I don't remember if we finished all the runs, e.g., magma?

\subsection{Usability Assessment}

% Scores! Assuming 1 == low, 2 == medium, 3 == high
% AWS Parallel Cluster (CPU, 7 and GPU, NA)  
% Microsoft CycleCloud (CPU, 15 and GPU, 15) 
% Google Compute Engine (CPU and GPU)
% Amazon EKS (CPU,9 and GPU, 11)
% Google GKE (CPU,7 and GPU, 7)
% Azure AKS (CPU, 15 and GPU, 12)

We assessed each cloud environment across the categories of setup, development, application setup, and manual intervention, for each assigning a perceived effort score of \emph{low}, \emph{medium}, or \emph{high} (Table \ref{table:environment-characteristics}). We were unable to make an assessment of \gls*{aws} ParallelCluster GPU due to an inability to do a custom build that required a combination of newer orchestration software with older drivers, reducing our assessment from 12 to 11 cloud environments.  
% NOTE FROM V: this is briefly mentioned in discussion
% On-premises \emph{low} ratings do not reflect the true effort it takes to maintain software or hardware on such a system. Administrative and support staff mask the difficulty of maintaining, optimizing, and developing the software and hardware for these systems.

% Specifically, the NVIDIA V100 GPUs needed older drivers than were provided by the default pcluster deployment tool. We were able to fix this first issue with a custom image build. However, when the cluster was provisioned, the nodes failed to to provision due to a software installation error. Allowing the build recipe to update packages resolved that issue, however we were then faced with needing to hard code a specific network configuration into the image. Given a 48 hour reservation and spending a large amount of time debugging, we could not make the time to test this step to see if the cluster would have come up. Arguably, if we had been able to get a small test cluster earlier in the month, a request that was not possible, we could have done this development work before the study and run the experiments.

\medskip
\noindent{\bf Accounts and Resources.} 
% Provisioning accounts took 6 months, and was completed before the start of the study in August 2024. Contracts and resource acquisition needed to be negotiated with each cloud separately. 
We ascribe a \emph{low} difficulty score to Azure and Google, as we had no issues with quotas or provisioning of GPUs. We ascribe a \emph{medium} score to for acquiring GPUs on \gls*{aws}. An initial, small reservation request was made at the beginning of August for prototyping. The reservation was never granted, meaning development and testing was not possible using GPUs. The reservation was pushed to a 48 hour block during the last week of the month. We had no issues with CPU quotas. 

\medskip
\noindent{\bf Setup.} 
ParallelCluster CPU (\emph{medium}) required a custom build and multi-step configuration. Each of the CycleCloud CPU and GPU setups (\emph{high}) took over a day to deploy and interfaces were challenging to use, often going out of sync with the Azure portal. % The interfaces were often out of sync or one did not detect changes in the other, and resource creation interfaces were often challenging to use. %, and CycleCloud sometimes did not detect our cloud account resources, cluster changes, or permissions. Procedure that worked previously would stop working. % We also found it to be the case that a change to the UI would not result in a change to the cluster, or vice versa. % The experience was, from a usability standpoint, immensely arduous.
% For Microsoft Azure interfaces that required selecting an instance type for a targeted zone, it often was the case that selecting the zone did not reveal the instances. Instead we had to deselect all zones to see instances that were given to under our quota. % Finally, when using the web interface to create resources, it was challenging to know when to select extra options. We generally recommend to others to always pursue the most programmatic means to deploy as possible to avoid user interfaces, which can rapidly change. In the case of our experiments, this was not always possible.
%Most interactions with Google Cloud could be programmatically orchestrated from the command line via the ``gcloud'' client, and using Terraform. 
The Compute Engine CPU and GPU setups were a \emph{medium} effort because we could not customize configuration files for Cluster Toolkit. % For \gls{gke}, drivers worked to detect all NVIDIA V100 devices.

CPU and GPU setups on Azure required multiple stages of commands to bring up clusters (\emph{medium}). For GPU driver provisioning, despite enabling health checks, we experienced an issue where a node consistently came up with 7/8 GPU on the 32 node cluster. We spent 20-30 minutes debugging and trying to request a new node.  Releasing the resource did not resolve the issue, as the node was re-allocated. Anticipating this issue, we had asked for quota of 33 nodes, and were able to bring up a final working node, and remove the erroneous one. This issue has been reported before \cite{lange-2023}.

During \gls*{aws} \gls*{eks} acquisition of GPU nodes an erroneously created placement group led to a partial instantiation of the cluster. Debugging and fixing of the issue required additional work and substantial cost to complete the setup (\emph{high}).

% Another point of confusion was mapping an instance family (e.g., Standard NDSv2 Family vCPUs) with a specific instance type (e.g., ND40rs\_v2) that only share two common letters, meaning that a search for the known instance name will not produce any results in the quota interface. 

% For the second example - the instance might be "ND40rs\_v2" but the family is "Standard NDSv2 Family vCPUs" with an additional "S" and different structure. Extra options like "No infrastructure redundancy required" also were needed, and not obvious to add.

% I'm not putting this because I think it was just me.
% Another issue I had with using the interface was that I never got any command line authentication to work. My workaround was using the cloud shell for everything, and (for example, for AKS) copying the kube-config from the cloud shell to my local machine.

\label{sec:usability-development}
\medskip
\noindent{\bf Development.} 
\gls*{aks} environments for CPU and GPU required a \emph{high} level of effort. We needed to create a custom container base that would support proprietary software (e.g., hpcx, hcoll, and sharp), and develop a custom daemonset to install InfiniBand drivers \cite{aks-infinifand-install}.

For \gls*{aws} \gls*{eks} (high), we discovered issues with an erroneously created placement group \cite{eksctl-bug-placement}, and a missing cleanup step that broke provisioning \cite{eksctl-issue}. We did a custom build of the tool to run the study. For the largest cluster size (256 nodes) we ran out of network prefixes for the \gls*{cni} and fixed the issue by patching the \gls*{cni} daemonset to allow for prefix delegation to increase the number of addresses available.
% an issue that presented only when pods were created. % \gls*{aws} support could not diagnose the problem because the cluster had been taken down. On the last day of the study we brought up the cluster once more to debug on our own, and 
% We were able to fix the issue by applying two environment variables to the \gls*{cni} daemonset, allowing for prefix delegation to increase the number of addresses available.

Finally, due to issues with GPUs and Slurm using Cluster Toolkit, we developed custom Terraform deployments for Flux Framework on Google Cloud for each of CPU and GPU (\emph{medium}). % We used a derivative \cite{vanessa_2024_14396613} of a recipe provided by Google \cite{google-scientific-computing-examples} that also is the basis for the Cluster Toolkit version. 

% And they fired Ward :(
% Note that this used to be in discussion. I'm removing it because I don't think we should be telling any clouds how to fix things. I think they can infer this and would guess this is possible anyway.
% For the specific issue related to CNI, a basic calculation might be done to calculate the number of addresses available as compared to an expected range to be requested based on cluster size, and the environment variables to fix the issue applied in an automated fashion. In the case that an error like this occurs, we suggest to AWS that new models of support might be warranted for cases like ours where it is costly to leave up a large cluster for debugging. The model should not require the user to pay for a large cluster to remain up for an unknown period of time waiting for cloud support to debug it.

\medskip
\noindent{\bf Application Setup.} 
In total there were 220 unique container builds, 114 tested, %(including architectures that segfaulted and needed rebuilds, see Section \ref{section:amg}), 
97 intended to be used, and ultimately only 74 used due to the \gls*{aws} ParallelCluster GPU environment not being deployed. The Azure container bases used across environments, for each of GPU and CPU (\emph{high}), were challenging to build.

% This was calculated from the counts of our actual data, what we actually used across the entire study:
% 4+6+11+2+6+7+10+6+3+6+6+2+5
% Out[1]: 74

% This is the count in the table, the number we prepared to use.
% Out[2]: 97

% When we programatically get all repos / tag under converged computing, we find 114 containers. (this includes ones we didn't use)

Working with \gls*{ucx} on Azure proved to be highly challenging. We did not find suggested practices, and found better performance on \gls*{aks} using \texttt{OMPI\_MCA\_btl=\^{}openib} running in unified mode (\texttt{UCX\_UNIFIED\_MODE=y}) and the transport set to infiniband (\texttt{UCX\_TLS=ib}). On CycleCloud the best performing transport was \texttt{UCX\_TLS=ud,shm,rc} for unreliable datagram, shared memory, and reliable connected \cite{ucx-parameters-nvidia}. On-premises environments were \emph{high} difficulty due to less control over the software environment. 

\label{sec:manual-intervention}
\medskip
\noindent{\bf Manual Intervention.} 
For \gls*{aks} (\emph{high}) operations asking for proximity placement groups would not complete for 100 nodes or more, and while we were able to manually create and scale up the cluster, the interface reported that ``Colocation status is currently unknown'' and we verified that only a subset of nodes were included. 
% . We were able to create a size 99 cluster, and manually increase the size to get around the error. After the study we tested providing an additional placement group flag, and while it did not return an error, in practice it created a node pool with only one group and a message in the interface that the ``Colocation status is currently unknown.'' We verified our second proximity placement group was not associated with any resource, suggesting only a subset of nodes were included.

CycleCloud required a \emph{high} level of effort to manage jobs. %, because the environment was unstable. 
Job submissions needed to be monitored to ensure jobs were not stalled due to what appeared to be issues with process management, module loading, Slurm, or the environment. % Often a cancellation of a running job was required with submission of a new one.

We ascribe \emph{medium} effort to all other Kubernetes environments because of the manual work required to deploy a cluster across nodes, and shell in to interact with the queue for each application. % for each application and shell in to connect to Flux to submit jobs is manually intensive. 
There is a trade-off between needing to shell in multiple times and the benefits of modularity. Modularity afforded by containers allows for differences in environment between applications.

On-premises runs required \emph{medium} effort for manual intervention, as jobs often would error, and had to be monitored and debugged. Often the runs were not successful due to a bad node.

\begin{table*}
\small
  \caption{Environment Usability - Assessment of Effort}
  \label{table:environment-characteristics}
  %\begin{tabular}{@{}p{2.9cm}lllllll@{}}
  \begin{tabularx}{0.75\textwidth}{llllllll}
      \toprule
      \multicolumn{2}{c}{Environment} & Accelerator? &  Setup & Development & \makecell{Application \\ Setup} & \makecell{Manual \\Intervention} & \\
    \midrule
   Amazon Web Services & ParallelCluster & CPU & \medium  & \low  & \low  & \low  &  \\ 
   Microsoft Azure &	CycleCloud &  CPU & \high & \low & \high & \high  &  \\ 
   Google Cloud &	Compute Engine &  CPU & \medium  & \medium  & \low  & \low  &  \\ 
   Microsoft Azure & CycleCloud &  GPU & \high & \low & \high & \high  &  \\ 
   Google Cloud &	Compute Engine &  GPU & \medium  & \medium & \low  & \low &  \\ 
        \hline
   Amazon Web Services & Kubernetes &  CPU& \low & \high & \low & \medium  &  \\ 
   Microsoft Azure & Kubernetes &  CPU& \medium  & \high & \high & \high   &  \\ 
   Google Cloud & Kubernetes &  CPU & \low  & \low  & \low  & \medium   &  \\ 
   Amazon Web Services 	& Kubernetes & GPU & \high & \high & \low  & \medium   &  \\ 
   Microsoft Azure & Kubernetes & GPU & \medium 	& \high & \high & \medium   &  \\ 
   Google Cloud & Kubernetes & GPU	& \low  & \low  & \low 	& \medium  &  \\ 
        \hline
   Institutional & On-premises & GPU 	& \low  & \low  & \high & \medium  &  \\ 
   Institutional & On-premises & CPU  	& \low  & \low  & \high & \medium  &  \\ 
  \bottomrule
\multicolumn{8}{l}{\footnotesize Qualitative usability assessment of different environments for procedure in each step. A value of ``low'' indicates that the procedure } \\
\multicolumn{8}{l}{\footnotesize required for the step worked after following available instructions, with minimal development or configuration needed. A value of  } \\
\multicolumn{8}{l}{\footnotesize  ``medium'' reflects facing unexpected issues to be resolved by debugging or development, and a value of ``high'' means that there was } \\
\multicolumn{8}{l}{\footnotesize  significant effort needed for components of the environment to function. } \\
\end{tabularx}
\end{table*}

\subsection{Networking}

We got \emph{COMPACT} placement on \gls*{gke} up to 128 nodes. % , and ran the size 256 study without it. 
We were not able to get any study size with \emph{COMPACT} placement for Compute Engine. \gls*{aks} had a proximity placement group that did not extend beyond 100 instances, as previously reported.

For Microsoft Azure, we measured a large \emph{hookup time} -- the time between when the job started and when the application would start running -- via LAMMPS, which reports application wall time that can be subtracted from the workload manager wrapper time. In the most extreme case for LAMMPS, we found the hookup time for sizes 4, 8, 16, and 32 GPU nodes to be approximately 43, 30, 20, and 10 seconds, respectively, while other clouds were consistently between 3-4 seconds across sizes. Interestingly, for the GPU setup the hookup time seemed to be slower with fewer nodes, a pattern that flipped for CPU runs, where the analogous environment for sizes 32, 64, 128, and 256 nodes had hookup times of approximately 50, 100, 200, and over 400 seconds for the same problem size. Other clouds consistently had times between 10-15 seconds across sizes, suggesting that scale was not a factor.

% https://github.com/converged-computing/performance-study/tree/main/analysis/lammps-reax#hookup-time-for-gpu-64-x-64-x-32

% Note that we can try to better mirror the structure of methods and results - the numbers likely don't line up now because all sections aren't shared. We can always remove numbers too.

\subsection{Applications}

We present \gls*{fom} values for applications in the sections below, highlighting again that Google Cloud instances had fewer CPU cores per node (56) than instances on AWS and Azure (96).

% Where applicable, we plot the speedup, calculated by dividing the \gls{fom} value at each test size (number of nodes or GPUs) by the median \gls{fom} at the smallest test size.
% is the relative scale of each subsequent cluster size (nodes or GPUs) as compared to the smallest. In practice, this means calculating the median value for the smallest size of each environment, and dividing the values at each subsequent test size by the median. %In this way we normalize by the median of the smallest test size. 

% Things we might want to write about?
% How we messed up running benchmarks (e.g., mixbench) in terms of our parameter selection
% Laghos that didn't seem to work in most places and we don't know why.

% Observation
% - AMG cpu Dane dominates, but Lassen GPU it sucks (tied for last)
% But for LAMMPS dominating for GPU and CPU
% Maybe memory bandwidth, host to GPU memory bandwidth? (could compare cuda stream tests on cloud vs. lassen). If it's related to memory bandiwdth we'd expect lassen to not do well on the stream test as compared to lassen.

% Not sure the information about the segfaults is necessary

\begin{figure*}[h!]
    \includegraphics[width=1.0\textwidth]{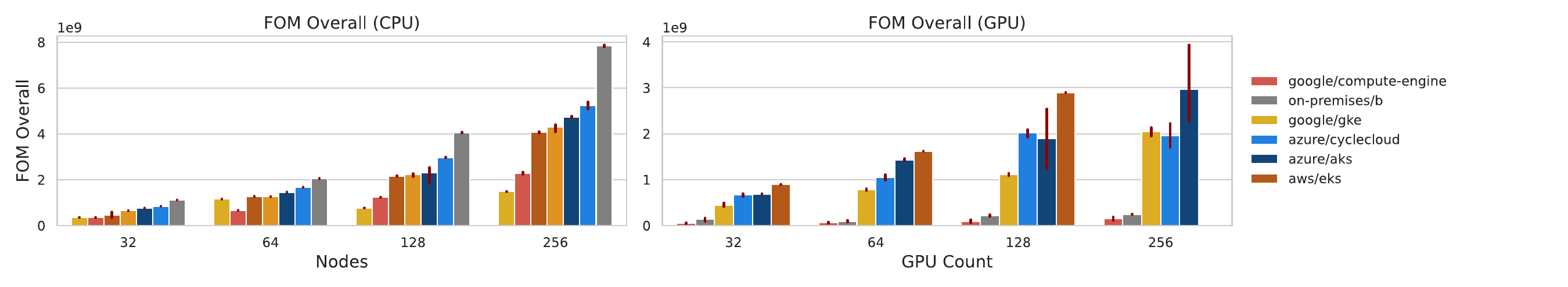}
  \caption{AMG2023 Overall FOM for each of CPU and GPU. Higher values are better. Cloud environments excelled for GPU runs, while on-premises had the highest FOMs for CPU. AMG2023 is weak scaled, so problem sizes are different.}
  \label{fig:amg}
\end{figure*}

\label{sec:amg2023-result}
\medskip
\noindent{\bf AMG2023.} 
 % The Google Cloud container required a rebuild due to segfault. The architecture generation deployed on the same instance type used to build the image was much newer than the one deployed at scale. 
 Results for CPU and GPU are in Figure~\ref{fig:amg}. Larger \gls*{fom}s indicate better performance. We note discrepancies between Compute Engine and \gls{gke}. The on-premises cluster \emph{B} (GPU) produced some of the lowest \gls*{fom}s across sizes, but cluster \emph{A} (CPU) produced the largest. We discovered an inconsistency in AMG process topology settings~\cite{amg2023-github} that affected the size 64 GPU tests. For \gls*{vm} environments, we used a topology of \texttt{-P 4 4 4}. We used a topology of \texttt{-P 8 4 2} for the Kubernetes environments. We compared these two topologies using a size 64 \gls*{gke} test to determine that \texttt{-P 8 4 2} results in about 10\% higher \gls*{fom} than \texttt{-P 4 4 4}. The results for \gls*{gke} at size 64 GPUs in Figure~\ref{fig:amg} are generated from \texttt{-P 8 4 2}.

\begin{figure}[b]
  \centering
  \includegraphics[width=\linewidth]{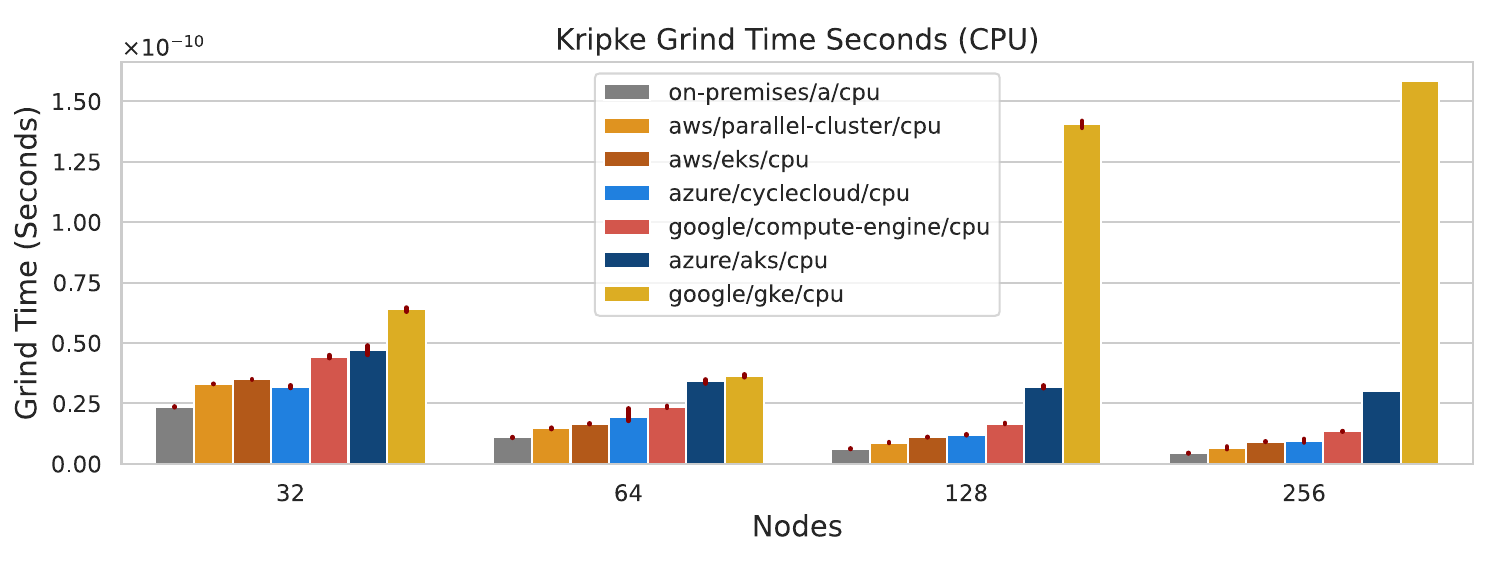}
  \caption{Kripke \emph{grind time} -- the amount of time to complete a unit of work for CPU environments. Lower values are better.}
  \label{fig:kripke-grind-time}
\end{figure}
% I'm not sure what description is (it doesn't seem to show up) but I get a warning without it.

% This was the case where skylake worked, so we assumed zen4 would, but then when we ran the study it segfaulted. We found it was using haswell. I don't know the degree to which we can relate this to the special bigint vs. mixedint flags - to get it to work I had to rebuild from scratch (not using spack) with the same flag pattern, and that worked, so arguably it is something about spack.

\label{sec:kripke-result}
\medskip
\noindent{\bf Kripke.} 
\gls*{aws} ParallelCluster had the lowest grind time for the largest three sizes (CPU), followed by \gls*{eks} and CycleCloud. While CPU and memory are likely factors, we believe network interconnect to be the strongest influence (Figure \ref{fig:kripke-grind-time}). We do not report GPU runs due to difficulties mapping processes to GPUs correctly. 

\begin{figure}[b]
    \includegraphics[width=0.48\textwidth]{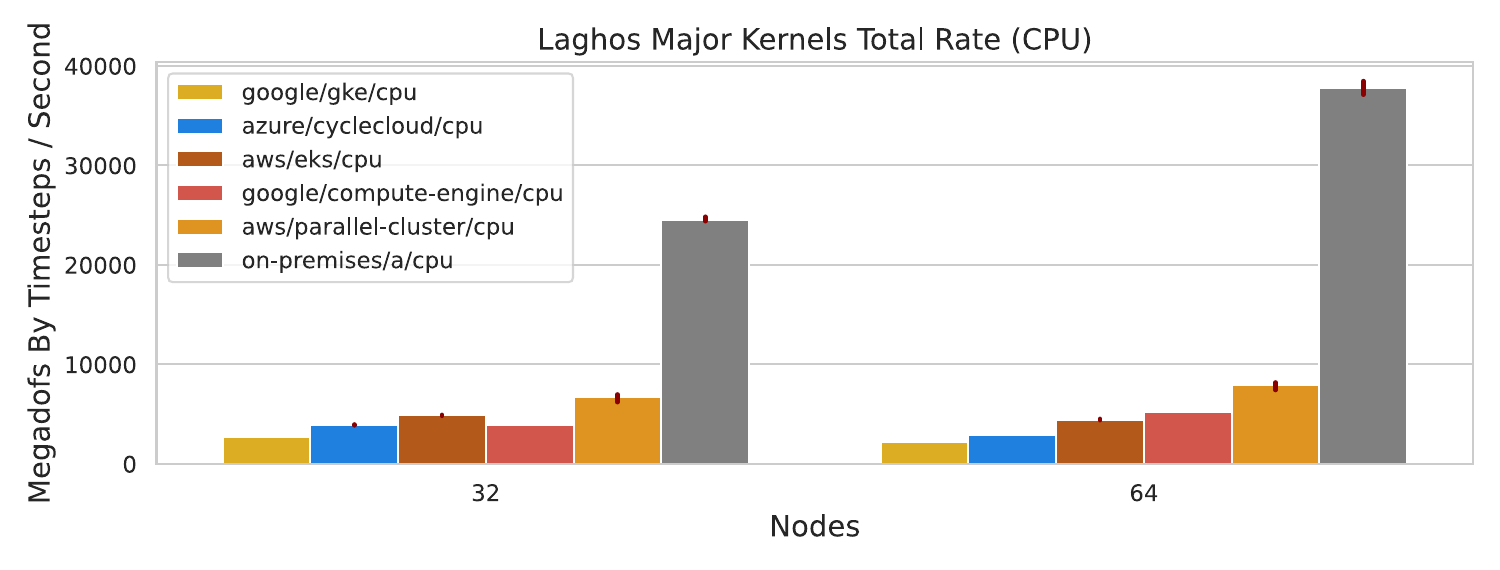}
  \caption{Laghos Major kernels total rate (megadofs $\times$ time steps / second) on CPU. Larger values are better. Laghos ran well on-premises HPC and was challenging to run and strong scale on cloud.}
  \label{fig:laghos}
\end{figure}

\label{sec:laghos-results}
\medskip
\noindent{\bf Laghos} 
only completed successfully on sizes 32 and 64 nodes (CPU) in all cloud environments except for \gls*{aws} ParallelCluster. Beyond 64 nodes % in all cloud environments except \gls*{aws} ParallelCluster, 
it experienced increasing slowdown that prevented runs from completing in under 15-20 minutes. % with the exception of \gls*{aws} ParallelCluster where it ran up to 128 nodes. % but we exclude that node size in Figure~\ref{fig:laghos}. % The on-premises runs on cluster \emph{A} experienced meshing errors at the 64-node size, but the runs still completed successfully. 
For on-premises runs, at 128 and 256 nodes Laghos experienced segmentation faults on cluster \emph{A}. The on-premises \gls*{fom} is one order of magnitude larger than the cloud environments, and exhibits a speedup of nearly 1.6 with lower variability in comparison with cloud (Figure~\ref{fig:laghos}). We were not able to build GPU containers due to a software conflict of two dependencies requiring different versions of CUDA. Due to the inability to scale, Laghos would be infeasible to run on any cloud. 

\begin{figure*}
    \includegraphics[width=1.0\textwidth]{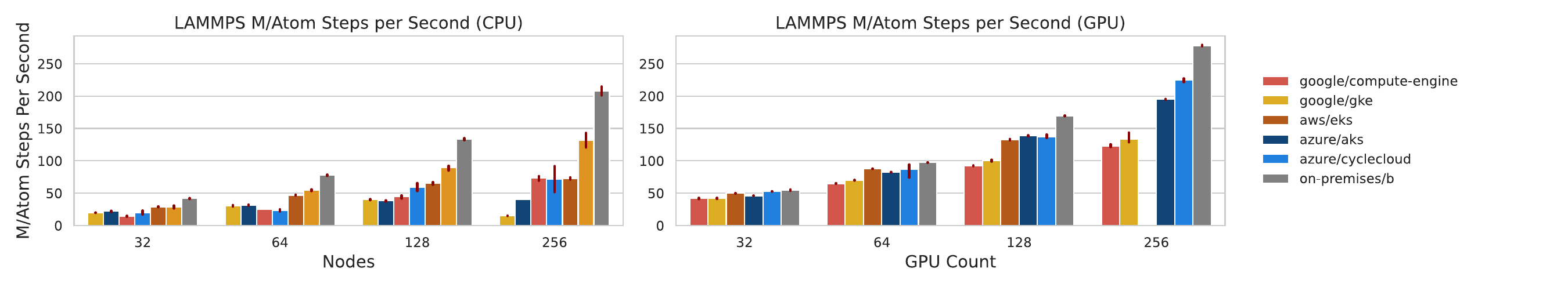}
  \caption{LAMMPS millions of atoms steps/second for each of CPU (problem size 64x64x32) and GPU (problem size 64x32x32). GPU runs were not possible on AWS Parallel Cluster, or for the largest size of AWS EKS.}
  \label{fig:lammps-matom}
\end{figure*}

\label{sec:lammps-reax-results}
\medskip
\noindent{\bf LAMMPS} 
 \gls*{fom} is Matom (million atom) steps per second, where larger values indicate more work performed in the same amount of time. Only one run was performed for \gls*{aks} CPU at size 256 due to long hookup time (8.82 minutes), and the largest size on \gls*{eks} was not possible due to inability to get GPUs. On \gls*{gke} CPU the runs exhibited an inflection point between 128 and 256 nodes where strong scaling stopped. On-premises clusters \emph{A} and \emph{B} produced larger \gls{fom}s in comparison to cloud environments (Figure~\ref{fig:lammps-matom}).

\label{section:mixbench}
\medskip
\noindent{\bf Mixbench} 
revealed an interesting setting of Error Correction Code (ECC). ECC is a GPU setting that typically defaults to \emph{On} and is necessary to maintain the data integrity for many scientific codes. Turning it on can reduce performance by up to 15\%, however it is typically the default setting to maintain data integrity \cite{copeland2009gpu}. All cloud GPU environments except Azure turned ECC on, meaning all GPUs were using error correction. Azure had a mixture of settings across environments, ranging from 12.5-25\% for \emph{Off} and 50-100\% for \emph{On}. This inconsistent setting warrants further investigation.
% The one environment that had 100\% of GPUs with ECC On was a small 4 node cluster on AKS. 

\begin{figure*}
  \includegraphics[width=1.0\textwidth]{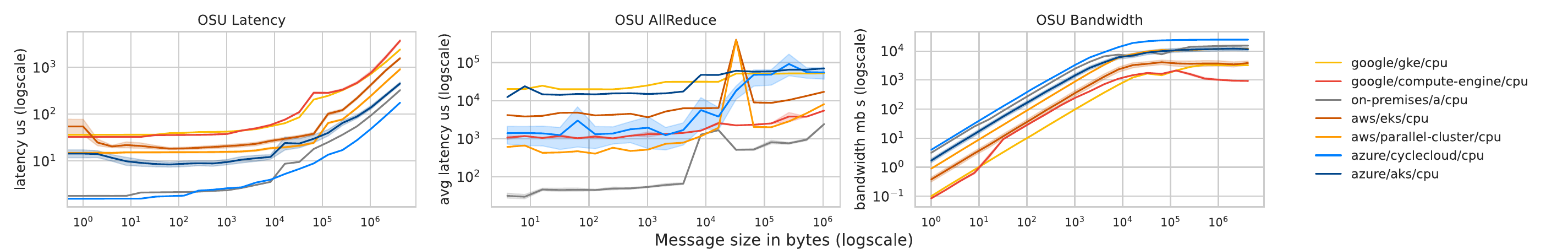}
  \caption{OSU Benchmark results for CPU cluster size 256 nodes for cloud and on-premises environments. Infiniband and on-premises low latency fabrics had the lowest latencies. CycleCloud had the highest variation within runs for OSU AllReduce.}
  \label{fig:osu}
\end{figure*}

\label{sec:osu-results}
\medskip
\noindent{\bf OSU Benchmarks} 
are shown in Figure~\ref{fig:osu}. GPU and CPU results were comparable, and so CPU are reported for the largest cluster size (256 nodes). For \gls*{eks} and \gls*{aks} environments, the point-to-point latency and bandwidth tests were run simultaneously on the same nodes, likely negatively impacting performance.  % We ran point-to-point benchmarks 28 times using a random sampling method described in Section~\ref{sec:applications}. AllReduce tests were repeated five times. 
For OSU latency, environments with InfiniBand fabrics (on-premises \emph{A}, and Azure CycleCloud) had the lowest latency values. The AllReduce test depicts a latency spike for both \gls*{aws} environments at a message size of 32,768 bytes. This is a known performance issue that has been addressed by a recent improvement \gls*{aws} made to OpenMPI AllReduce~\cite{aws-openmpi}. The highest bandwidth was seen for Azure CycleCloud. Smaller cluster sizes (not shown) showed similar patterns and relationships between environments~\cite{vanessa_2024_14396613}.

% AWS PC CPU is the next lowest for osu all reduce
% AWS EKS GPU and Lassen are lowest for GPU all reduce
% Note: we are not including this for now, missing on premises data

\begin{figure*}[h!]
    \includegraphics[width=1.0\textwidth]{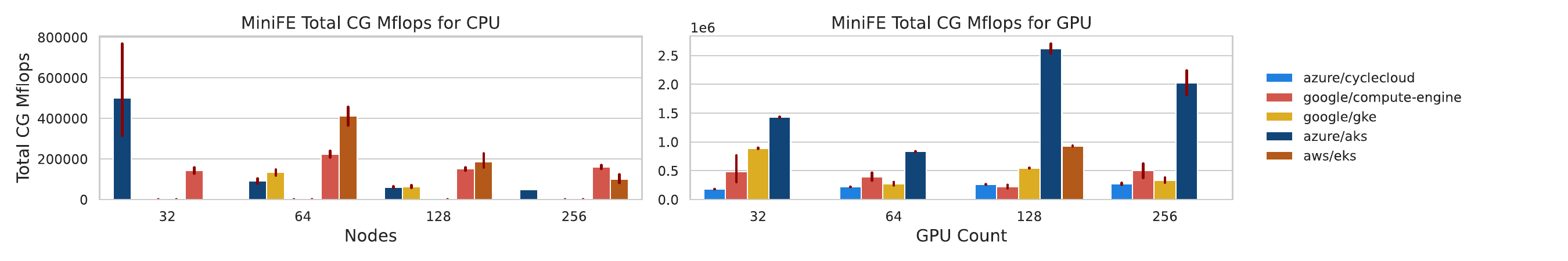}
  \caption{MiniFE (Mini Finite Element Conjugate Gradient) millions of floating-point operations per second. Higher values are better.}
  \label{fig:minife}
\end{figure*}

\label{section:miniFE}
\medskip
\noindent{\bf MiniFE} 
results (CPU and GPU) are in Figure \ref{fig:minife}, showing millions of floating-point operations per second. Higher values are better. \gls*{aks} exhibited the best performance for GPU, and for size 32 CPU. The results across both CPU and GPU indicate inconsistent and inverse scalability.
We hypothesize that the inconsistency and overall inverse scaling is due to network limitations, or GPU misconfiguration. Further investigation is needed. For on-premises studies, partial output was saved and we are not able to report the result. \textbf{}

% TODO what do we want to say?

% Here is the two processor one:
% Azure AKS, size 32, node-19
% https://github.com/converged-computing/supermarket-fish-problem/tree/35792e2a790c3e85df8b0ae6bcd17f844c9a91c4/data/azure/aks/gpu/32/node-19/raw

\label{sec:mt-gemm-result}
\medskip
\noindent{\bf MT-GEMM.} 
The performance of the matrix-matrix multiplication performed by MT-GEMM is measured in GFLOPs. The GPU tests exhibit strong scalability across GPU sizes, with Google Compute Engine, Azure AKS, and GKE exhibiting similar performance. % CPU runs were not used due to a hard coded problem size that made the problem size per rank small enough to introduce a communication bottleneck.
We omit the results of the CPU runs. MT-GEMM hard codes the global problem size in its source code, and the per-rank problem size is small even at the smallest CPU node count. At that initial per-rank problem size all CPU environments are already communication bound, and GFLOPs decreased at each larger node count.

%\begin{figure*}
%    \includesvg[width=1.0\textwidth]{images/apps/mtgemm-cpu-gpu.svg}
%  \caption{MT-GEMM Giga Floating-Point Operations (GFLops) per Second that represents efficiency of matrix multiplication. 1 GFlops is a billion calculations each second. Compute Engine is fastest for CPU, and tied with Azure AKS and GKE for GPU. Larger values are better.}
%  \label{fig:mt-gemm}
%\end{figure*}

\begin{figure}
    \includegraphics[width=0.5\textwidth]{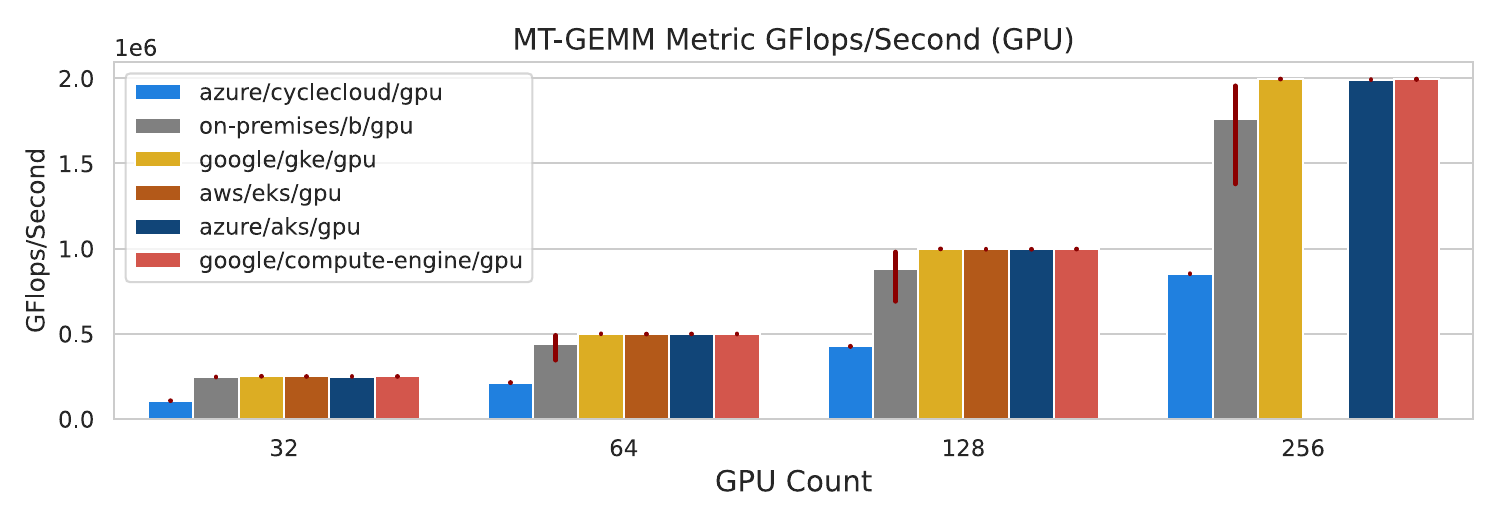}
  \caption{MT-GEMM Giga Floating-Point Operations (GFLops) per second showing efficiency of matrix multiplication (GPU).  AWS Parallel Cluster was not run. Larger values are better.}
  \label{fig:mt-gemm}
\end{figure}

\label{sec:single-node-results}
\medskip
\noindent{\bf Single Node Benchmark} 
We found consistent machines with one exception on \gls*{aks}, where there was one instance that only reported two processors across collection mechanisms. We call this the supermarket fish problem \cite{supermarket-fish-problem}, the idea being analogous to buying generic white fish at the supermarket (an instance type) but in reality receiving any one of a multitude of species (architecture generations or other differences). % Google Cloud had consistent clock speeds, with all nodes reporting the same value, while \gls*{aws} ranged from less than 1k GHz to 3.5k GHz. The scaling governor did not explain the differences \cite{scaling-governor}. The observation is not yet explained. % The other case of this problem occurred with AMG, but not during a single experiment, but rather, between them at multiple scales, discussed next.

\label{section:stream-results}
\medskip
\noindent{\bf Stream} 
was run with several configurations \cite{vanessa_2024_14396613}. To measure memory bandwidth for CPU, we used the Stream Triad Kernel and found relatively comparable performance with high variation across \gls*{gke} (6800.9 $\pm~2402.29$), Compute Engine (6239.35 $\pm~2326.1$), \gls*{eks} (3013.23 $\pm~880.3$) and \gls*{aks} (2579.5 $\pm~907.58$), all in GB/s, for a size 64 cluster. For the GPU Triad we report results in GB/s from the largest size 32 cluster. We found comparable results across sizes for \gls*{gke} (782.91 $\pm~0.72$), Compute Engine (783.3 $\pm~0.73$), \gls*{aks} (748.54 $\pm~4.63$), on-premises \emph{B} (782.52 $\pm~ 0.96$), and Azure CycleCloud (748.54 $\pm~4.63$).

\begin{figure}
  \centering
  \includegraphics[width=0.5\textwidth]{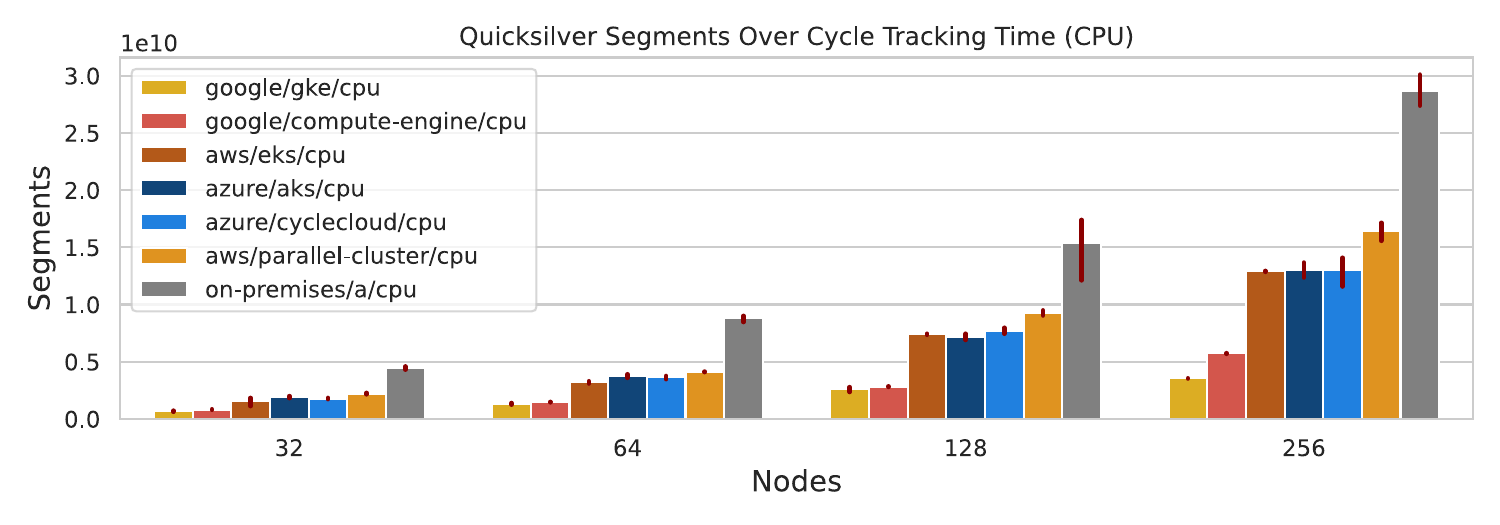}
  \caption{Quicksilver number of segments over cycle tracking time. Larger values are better.}
  \label{fig:quicksilver}
\end{figure}

\label{section:quicksilver}
\medskip
\noindent{\bf Quicksilver} 
had the highest number of segments over cycle tracking time for \gls*{aws} setups, followed by Azure (Figure \ref{fig:quicksilver}).  Runs on GPU did not finish within the allocated time dictated by our budget. We noticed poor GPU utilization and that half of processes were pinned to GPU 0, a likely erroneous build or runtime misconfiguration that would require further debugging.

\subsection{Costs}

The study spent \$31,056 on Azure, \$31,565 on \gls*{aws}, and \$26,482 on Google Cloud. We were under-budget because the GPU experiments on \gls*{aws} ParallelCluster were not possible to run, and the Compute Engine GPU environment on Google Cloud was funded by credits. Further, we estimated Kubernetes costs assuming running all sizes on a single cluster of the largest size, and ultimately created separate clusters of each size. While we are not able to fairly compare costs between clouds from our billing statements due to human variation, we can compare costs for a representative app, AMG2023, that weak scaled well. The total cost by environment accounting for execution time, cluster size, and instance cost is shown in Table \ref{table:costs}. 

% Note that this accounts for each runtime and size.
% azure/cyclecloud/gpu          84.081167
% aws/parallel-cluster/cpu         93.312
% aws/eks/gpu                   96.525686
% azure/aks/gpu                 97.891336
% google/gke/gpu               133.417111
% google/compute-engine/gpu    135.256841
% aws/eks/cpu                  266.731258
% azure/cyclecloud/cpu            348.832
% azure/aks/cpu                358.192452
% google/compute-engine/cpu    396.780725
% google/gke/cpu               487.407771
% I don't think we should put that we only had a month, were under staffed, etc. It sounds like a lame excuse and I think in practice isn't the main reason we were under budget.

\begin{table}
\small
  \caption{AMG2023 Total Costs By Environment}
  \label{table:costs}
  \begin{tabular}{@{}p{2.9cm}llll@{}}
      \toprule
      Environment & Accelerator? & Cost/Hr & Total Cost \\
    \midrule
    % GPU Azure CycleCloud       & LAMMPS & \$22.03 & \$84.08 \\
    % CPU ParallelCluster        & LAMMPS & \$2.88 & \$93.31 \\
    % GPU \gls*{aws} \gls*{eks}  & LAMMPS & \$34.33 & \$96.53 \\
    % GPU Azure \gls*{aks}       & LAMMPS & \$22.03 & \$97.89 \\
    % GPU Google \gls{gke}       & LAMMPS & \$23.36 & \$133.42 \\
    % GPU Compute Engine         & LAMMPS & \$23.36 & \$135.26 \\
    % CPU \gls*{aws} \gls*{eks}  & LAMMPS & \$2.88 & \$266.73 \\
    % CPU Azure CycleCloud       & LAMMPS & \$3.60 & \$348.83 \\
    % CPU Azure \gls*{aks}       & LAMMPS & \$3.60 & \$358.19 \\
    % CPU Compute Engine         & LAMMPS & \$5.06 & \$396.78 \\
    % CPU Google \gls*{gke}      & LAMMPS & \$5.06 & \$487.41 \\
    Azure \gls*{aks}           & GPU & \$3.60 & \$13.82 \\
    Azure CycleCloud       & GPU & \$22.03 & \$23.01 \\
    Google \gls{gke}       & GPU & \$23.36 & \$13.82 \\
    \gls*{aws} \gls*{eks}  & GPU & \$34.33 & \$33.74 \\
    Azure CycleCloud       & CPU & \$3.60 & \$137.79 \\
    ParallelCluster        & CPU & \$2.88 & \$187.39 \\
    \gls*{aws} \gls*{eks}  & CPU & \$2.88 & \$263.75 \\
    Compute Engine         & GPU & \$23.36 & \$266.84 \\
    Azure \gls*{aks}       & CPU & \$3.60 & \$309.07 \\
    Google \gls*{gke}      & CPU & \$5.06 & \$477.46 \\
    Compute Engine         & CPU & \$5.06 & \$591.54 \\
% AMG
% azure/aks/gpu                 13.817773
% azure/cyclecloud/gpu          23.009111
% google/gke/gpu                30.865626
% aws/eks/gpu                   33.744743
% azure/cyclecloud/cpu            137.792
% aws/parallel-cluster/cpu        187.392
% aws/eks/cpu                  263.748075
% google/compute-engine/gpu     266.84218
% azure/aks/cpu                309.074591
% google/gke/cpu               477.464244
% google/compute-engine/cpu    591.538002

  \bottomrule
\multicolumn{4}{l}{\footnotesize Costs for representative app, AMG2023, that weak scaled well.} \\
\multicolumn{4}{l}{\footnotesize Total is a summation across iterations, accounting for nodes and instance cost.} \\
\end{tabular}
\end{table}

%    instance_costs = {
%       'google/gke/cpu': 5.06,
%       'google/gke/gpu': 23.36,
%       'google/compute-engine/cpu': 5.06,
%       'google/compute-engine/gpu': 23.36,
%       'aws/eks/cpu': 2.88,
%       'aws/eks/gpu': 34.33,
%       'on-premises/dane/cpu': None,
%       'on-premises/lassen/gpu': None,
%       'aws/parallel-cluster/cpu': 2.88,
%       'azure/cyclecloud/cpu': 3.60,
%       'azure/cyclecloud/gpu': 22.03,
%       'azure/aks/cpu': 3.60,
%       'azure/aks/gpu': 22.03, 
%    }

% AMG
% azure/aks/gpu                 13.817773
% azure/cyclecloud/gpu          23.009111
% google/gke/gpu                30.865626
% aws/eks/gpu                   33.744743
% azure/cyclecloud/cpu            137.792
% aws/parallel-cluster/cpu        187.392
% aws/eks/cpu                  263.748075
% google/compute-engine/gpu     266.84218
% azure/aks/cpu                309.074591
% google/gke/cpu               477.464244
% google/compute-engine/cpu    591.538002

% LAMMPS
% azure/cyclecloud/gpu          84.081167
% aws/parallel-cluster/cpu         93.312
% aws/eks/gpu                   96.525686
% azure/aks/gpu                 97.891336
% google/gke/gpu               133.417111
% google/compute-engine/gpu    135.256841
% aws/eks/cpu                  266.731258
% azure/cyclecloud/cpu            348.832
% azure/aks/cpu                358.192452
% google/compute-engine/cpu    396.780725
% google/gke/cpu               487.407771

\section{Discussion}
\label{sec:discussion}

We completed a cross-cloud performance study of 11 applications on 11 cloud environments and two on-premises clusters and discussed the results with three cloud providers to promote collaboration. In an ideal study, we would have been able to get the latest generation GPU across clouds in the capacity needed, and use the latest technologies for networking and performance that they come with. We made a best effort to choose optimally given limited GPU availability, the desired scale, and a preference for common hardware across clouds. % For this reason, while we are measuring what is traditionally performance, we recognize that we missed improvements in performance by way of this choice. % In addition, we made a choice of breadth over depth for application selection, choosing to get 12 applications built and running across 14 environments over fewer applications working to some hypothetical optimum. This choice reflects a trade-off between performance and portability. Moving from HPC systems to cloud often comes with a transition from spending months of time to fine-tune application performance to exact hardware, to spending less time to run more generic container builds on nodes that are not entirely under user control. 
We note that a study of this scale required knowledge of the computing roles of developer, administrator, and end user. % The following sections highlight discussion on to performance and usability for each cloud, and identify future work.

\subsection{Insights}

With resource contention, a changing economic landscape, and often limited access to compute environments, scientists and \gls*{hpc} practitioners can maximize resiliency to change by embracing portability. While we do not see access to \gls*{hpc} going away, we want to be prepared for a future when our scientific work increasingly needs to run on imperfect, changing environments. The doctrine of ``best practices'' that encompass specific building and strict configuration of applications is replaced with best practices that are focused around flexibility and adaptation with reasonable performance. This study was an exercise in testing this mindset and method of work. It rests on a non-traditional premise: the end goal is not to directly make claims about performance, but to reflect on and advise for the experience of working in ephemeral environments. We test scaling of \gls*{hpc} apps to understand what kinds are strongest contenders for cloud orchestration, and we test usability to provide early guidance on challenges and opportunities for working in the space. 

\medskip
\noindent{\bf Hypervisor environments can be suitable to some \gls*{hpc}} \\
From our work we believe that cloud environments equipped with OS-bypass or \gls*{rdma}-based networks can be well-suited to running small to medium scaled HPC apps, in the span of tens to a few hundred CPU nodes. For the \gls*{hpc} user that is  interested in apps that strong scale like LAMMPS or weak scale like AMG2023, there is precedence to run them in cloud, and we are confident in the configurations we used. Larger runs are not likely to be possible in the near future without substantial monetary investment and special circumstances \cite{Liu2020-uk}. As a result of resource contention due to \gls*{ai}/\gls*{ml}, GPUs can be used in smaller quantities, and older generations can be more amenable to access at the cost of potentially deprecated support. Planning for execution, costs, and error resolution in a timely manner is a new mindset that must become a standard.

% Interesting that GPUs work better, more incentive to 

\medskip
\noindent{\bf Portability is a new dimension of performance} \\
% The ability to move flexibly across environments requires increased portability and generalization, which often comes at the cost of performance. Spending a significant amount of time optimizing applications for a single environment might not be possible or desirable if applications move frequently. 
Portability presents a new optimization problem: time spent optimizing performance on a single platform may be better spent on increasing portability or optimizing less on more platforms. Portability increases the size of the pool of suitable resources, enabling the capability to decide where to run based on when resources are available and which are best or most cost-effective. % An application run at some scale that is ``optimally good'' much later may not be as ideal as ``good enough'' at a much earlier time-point. 
% In this sense portability becomes a dimension of performance. 
Containers or binary caches provided by package managers are important for this goal, as system-optimized binaries are not portable.
Cloud environments will continue to quickly evolve, and to successfully harness them we must adopt a strategy to move equivalently. % However, there are caveats. Laghos exemplifies an application that did not easily port to cloud, and for which an initial run at any scale would not be feasible. More time was needed to optimize its execution.

\medskip
\noindent{\bf Environments are ephemeral and inconsistent by default} \\
The design of our study reflects an increasingly present reality for the scientific community: having an ability to quickly utilize environments that become available will be a strategic benefit to support timely execution of workflows. The traditional practice of build, test, and run in one persistent location that will be maintained by a team of administrators for years is replaced with the need to independently do orchestrated, complex work on transient environments that change quickly, having components that are opaque or poorly documented. Every cloud provides different interfaces, cost models, and tooling to deploy resources. Testing for a resource type or scale is more often not possible, and support often falls on the scientist if they are not a top-priority customer. A mindset of slow, careful work is replaced with a requirement of resiliency and an ability to respond quickly to the unexpected. % If teams remain small or further shrink due to lack of scientific funding, the task falls on the few. In our case, more than 50 experiments, each lasting a few hours, were run in under two weeks, by a few individuals. %  with a team of five, with two individuals doing the majority of work. 
% We often consider whether applications or resources scale, but we do not equivalently consider people. % Running scientific workflows in cloud is entirely different than \gls*{hpc} as every component is a moving target. 

\medskip
\noindent{\bf Scientific work will be carried out where it is not intended} \\
%Given the pace of change of cloud, the lack of control or transparency of underlying systems, and a likely future where the scientific community will not have the financial backing to be considered a first-tier customer, 
We cannot assume there will soon be a reality where cloud environments have components that are optimized for scientific use cases. The pace of change that is forced on relatively small teams might mean that we continue to be responsible for learning and executing the setup of infrastructure, software, and orchestration. The perception that \gls*{hpc} centers are more unique and brittle is false based on our observations. In our experience, every cloud environment has nuances with respect to network, compute, and tooling. Furthermore, the high rate of change of cloud infrastructures translates to rapidly changing brittleness. The challenge has not changed, only the context. It is increasingly important to identify shared goals between science and industry to ensure the technology space continues to evolve in a way that can support both.

\label{section:performance-discussion}
\medskip
\noindent{\bf Common tooling promises to empower small teams} \\
A small team can be empowered by standards, and community convergence on shared tooling such as Kubernetes, which  already has adoption in the broader computing community. We believe that if the scientific community can also collaborate on Kubernetes, it will be further used for complex scientific workflows. While networking can be a performance bottleneck, the use of OS-bypass mechanisms and \gls*{rdma} should provide comparable performance of Kubernetes to a cluster of \gls*{vm}s. In this study we did not consistently measure comparable performance to \gls*{vm} configurations, nor did we consistently measure performance degradations such as those reported in \cite{Beltre2019-mr}. Given the importance of this standard infrastructure and means to deploy applications and services, we anticipate future work to better understand the performance delta.

\medskip
\noindent{\bf Extended cost and scheduling models are needed} \\
Cost is an important facet of usability, as an experiment that is not cost-effective may not be tenable.  Resource selection considerations that account for hardware age, capacity, and cloud support are often in conflict, presenting challenges for the researcher. It is frequently impossible to obtain newer GPUs at the scale that is needed for performance testing or scientific investigation, regardless of ability to pay.  If the researcher uses older GPUs, the software and drivers are more likely not to work or be unsupported, as we discovered in our study. In \gls*{hpc}, a challenge is not obtaining resources, but allocating them with high utilization. %Queue time is often long because utilization is high. 
Cloud could address the resource availability challenge by providing a queuing and scheduling system with estimated job start times based on resource availability, similar to \gls*{hpc}. Capacity blocks from AWS \cite{Amazon-Web-Services2023-capacity-blocks} or Google's Dynamic Resource Scheduler \cite{Lohmeyer2023-hk} are improvements, but are limited in terms of resource type and the quantity that can be reserved.

We attempted to recreate a size 256 cluster on \gls*{eks} and reproduced a finding from \cite{lange-2023} for which the total number of nodes were never provisioned, and we were charged upwards of 2.5k waiting for nodes. An improved model for provisioning is needed here. %The work from Lange e.g., led to resolution of this issue for ParallelCluster, however the problem still persists on \gls*{eks}. 

\medskip
\noindent{\bf Auto-scaling should be used carefully}

Auto-scaling is most useful for running batches of infrequent work. A small head node can remain up that scales when needed, with the goal of minimizing scaling operations and total time of nodes going up and down relative to the work. Anything that needs more regularly changing sizes we suggest to the reader to use Kubernetes. % Most \gls*{hpc} clusters do not support autoscaling, and workflow tools are relied on for orchestration.
For cases when experiment runs are well defined, a strategy of bringing up static clusters of exactly the sizes needed can avoid costs incurred waiting for resources ~\cite{hossen-2024}. % On our \gls*{vm} setups, instances would come up in about 3-4 minutes for CPU, or 7-8 minutes for GPU. We waited up to 20 minutes in the Azure environment % due to issues with autoscaling, provisioning nodes, and responding to failed jobs. 

\subsection{Suggested Practices}

\textit{\textbf{Containers}} are an important tool to enable portability, and are already used by approximately 80\% of the \gls*{hpc} community at least once a week \cite{hpc-containers-survey}. Our study demonstrated the remaining challenges of software and driver compatibility by building more than 220 containers for 12 environments. % This need primarily came down to differences in networking and system drivers, and fixing bugs that were discovered after initial builds. 
Improved interfaces to adapt to different devices and provide compatibility metadata \cite{oci-compatibility-working-group} will be increasingly important. 
% We recommended building Docker \cite{Merkel2014-da} containers, and pushing to a registry. 
Containers can be readily used with Kubernetes or \gls*{vm}s using Singularity \cite{Kurtzer2017-xj}. % This does, however, require building the initial Docker container with the knowledge that assets in root's home will not be accessible, and most locations will be read-only. 
For setups with a shared filesystem that dynamically add worker nodes, we suggest pulling containers once before spawning worker nodes. 

%\paragraph{Contract and Policy}

%One of the most challenging facets of the study was not related to the infrastructure or applications at all, but getting permission and the proper funding vehicles in place to support it. While these vary by institution, routing funding from an institution to not one, but three clouds, and often needing to find middle-men to handle the process, took upwards of 6 months to do, and almost went over the deadline when the funding would expire. In the case that a middle-man vendor is the only or best means to pursue, there are often overhead costs to additionally add to a plan.

% This is also a bit redundant with the methods section - let's discuss / consolidate.

% A cloud account can be fairly easy to setup, requiring often only institutional permission and a credit card. However, 

\textit{\textbf{Resources}} must be considered in the context of availability and acquisition strategies. % Preparing to use resources can take careful planning and time due to quota and resource requests and reservations. % An initial request is needed to ask for a higher quota or upper limit of a particular resource type (e.g., vCPU, GPU, or storage), and this can take anywhere from a day to a few weeks if the cloud does not have capacity. 
For some clouds, receiving quota is a confident assurance that a resource quantity is available, while for others, obtaining quota does guarantee that provisioning will be successful. For our study, getting GPU instances proved to be challenging, % While Azure and Google Cloud granted the GPUs needed after brief discussion of types and zones, \gls*{aws} required a request for a reservation early in the month of August, and then being 
and required us to be ``on call'' and ready for GPU capacity during off-hours. % for when a set would be available, and then to accept or reject a reservation, which could be offered during off-hours. % Largely, given the scarcity of the resources, 
Resource scarcity makes it less likely to have control over when resources are granted. Researchers need to be flexible.

%Choosing a GPU model and generation was a challenge in and of itself. % For our particular study, we wanted to get comparable architectures and hardware (e.g., GPU) across clouds, and this left us with not many options. We were able to determine that the NVIDIA V100s were available across clouds (and for our on-premises clusters). Using older GPUs meant that older software or drivers would likely not still be available everywhere. It is a challenging situation, either way, because 
% Researchers can't get newer GPUs that are in demand, but older GPUs (that possibly have a better price point) will be taken up because of exactly that. 

% This point is redundant (I say it elsewhere) so let's decide if we want to say it just once, and where.

\textit{\textbf{Development work}}. Many of our experiments would not have been possible without considerable debugging or development. % This process included creating installers for Infiniband drivers, common container bases for Azure, and fixing bugs with eksctl and \gls*{cni} for \gls*{eks}. 
Cloud providers should note that many scientific code teams may not have developers that can do this work. Many of the \gls*{vm} solutions that readily deploy a workload manager like Slurm are just starting points for what a user might ultimately need, and still require expertise in system administration and development.

% Debugging also included spur of the moment investigation of nodes that came up without the correct total number of GPU, and application performance that intuitively seemed against expectation. These tasks require expertise that spans cloud, HPC, and the applications themselves. % For our study, this happened several times, and led to several pointed re-runs of applications to resolve issues. We also made mistakes -- running an application with one setup, and then later in the study choosing a different strategy, and redoing runs. This expectation of mistake, which is reasonable for people to make, might suggest planning to have extra funds and time to redo work or debug. 
% Finally, we want to point out that some of our development was not required, but resulted from having a limited amount of time to complete the work. For issues with configuration, opening issues or scheduling meetings could have helped, but we didn't have the time.

Recognizing the lack of updated \gls*{vm}s and base containers for the larger \gls*{hpc} community to use on Azure, following the study we developed new \gls*{vm}s and matching containers on Ubuntu 24.04 with the latest drivers. Instead of using proprietary \gls*{mpi} and other associated software, we used an entirely open stack.

% Finally, we point out that development work required for on-premises application builds is often substantial because we do not control the environment.
%- in development work can also mention on premises 

% We can cite this when not blind
% https://github.com/converged-computing/flux-tutorials/tree/main/tutorial/azure,

\textit{\textbf{Cost Estimation}}. % is important, and is based on having an expectation about application runtimes with respect to the cost of resources. % A spreadsheet with applications, costs per hour, estimated run times, and iterations is a simple strategy to accomplish this goal. 
We advise adding a buffer for unexpected costs. % Depending on the cloud, the pricing models can vary, with some clouds offering prices for the entire node, and others requiring assembly of the price in cost units (e.g., memory, CPU, network). In the case of unit pricing or more convoluted pricing APIs, it can be challenging to estimate costs. % In our case, adding a bump to a tier of networking was hard to estimate the cost for, but we could estimate it being an additional 1/3 of the total compute cost by looking at historical experiment data. 
When feasible, we recommend employing test clusters to prepare experiments and test configurations. % For highly contended resources like GPU, getting a test environment can be challenging. Ideally research credits can be used, but if not, 
% Due to GPU availability, we were only able to test GPUs on one cloud before the study.
Clouds exhibit different cost reporting lag times, where usage and charging data may not be reported until the next day. Operating on a cloud environment with a one-day reporting delay warrants careful planning and pauses between experiments to ensure that costs are meeting expectations. An experiment can be adjusted to reduce cost if needed, but it is very difficult to fix overspending retroactively. The trade-off between node cost and execution time should be studied by benchmarking and scalability testing. For our representative application, AMG2023, the GPU runs were significantly cheaper despite the more expensive instance type. % For this reason, ensuring there is buffer time of a day between large runs is a good strategy. 

\subsection{Future Work}

Workload managers provide similar but subtly different interfaces for job submission, and having different managers across our environments added an additional layer of complexity. We anticipate future work to better package \gls*{hpc} workflows for Kubernetes, and better integrate traditional \gls*{hpc} schedulers with converged environments. % Manual orchestration and submission should not remain an enduring, arduous task.  % In this task we also were able to do an assessment of usability, and liked both Slurm and Flux for the submit interfaces. We found that only Flux was able to provide easy programmatic access to JSON formatted job information, events (with timestamps), and output, and for the other workload managers, we needed to wrap our application runs with timestamps to get a similar wrapping time. In the case that a run failed, we could not collect the final timestamp. % and could not assess the run duration or other events.

% Improvements to Google Cloud might resolve inconsistencies between Compute Engine and \gls*{gke}, including using instance types that provide the new Titanium micro-controllers and adapters \cite{google-titanium}. Google has launched a new community of practice, and several products that are oriented for efficient, scaled computation \cite{Ma-Weaver2024-en}.

We plan to study the long hookup times for InfiniBand on Azure, and to better understand the performance delta between Kubernetes and \gls*{vm}s.
% I don't remember if Azure offered this - I don't think so. We'd have to likely self fund. Not sure if we should exclude them like this (they were very positive about working with us, so I think we might just say all of them). 
Follow-up work is already underway to test newer hardware and environment offerings with \gls*{aws}. We look forward to sharing the work.

\section{Conclusion}

This cross-cloud study assessed the usability of cloud environments and performance of a suite of \gls*{hpc} applications on up to 256 nodes with 28,672 CPUs and 256 GPUs with \gls*{vm} environments and Kubernetes. The work generated collaborative discussion with cloud vendors to improve their environments for \gls*{hpc}. Our observations provide a starting point to identify causal analysis of issues with different setups and focused follow-up investigations. % Observed performance differences between Kubernetes and \gls*{vm} environments do not immediately suggest causes, which points to fruitful future work, as Kubernetes enables highly desirable capabilities for on-premises and cloud deployments.
Our study emphasizes that as computing environments become increasingly converged and complex, portability is now a dimension of performance. A traditional \gls*{hpc} strategy of optimizing the performance of an application for a small number of environments will be suboptimal as the requirements for portability increase. Although the pursuit of performance is still worthwhile and relevant, acknowledging that a greater understanding of the trade-off between portability and optimization is warranted. The contention for GPU resources may continue into the future, in which case sophisticated scheduling models and resource acquisition strategies will also be needed. %  jointly with application portability. %This is currently a blind spot for \gls{hpc} that we want to shed light on.

% Performance perfectionists - become obsessive about it and get into a competition with a number.

We encourage others in the \gls*{hpc} community to collaborate with cloud providers. We are currently working with individual clouds to address the issues that we discovered, and look forward to the follow-up studies our work inaugurates. There is no one-size-fits-all answer to the question of using cloud or HPC, and the adversarial dichotomy serves no benefit -- a dated anachronism in an ecosystem where the previously disparate technology spaces are becoming more seamless and integrated.

% I'm ded. Sleeps.

\begin{acks}
We want to thank Heidi Poxon and Evan Bollig (AWS), Bill Magro, Ward Harold, Kevin Jameson, and Steven Boesel (Google), and Andy Howard, Andrew Jones, Jerry Morey, Brandon Thompson (Microsoft) for their support and advice before, during, and after the study. We would like to also thank Steven Lopez (LLNL) that made much of the account setup possible.

This work was performed under the auspices of the U.S. Department of Energy by Lawrence Livermore National Laboratory under Contract DE-AC52-07NA27344 and was supported by the LLNL-LDRD Program under Projects No. 22-ERD-041 and 24-SI-005 (LLNL-CONF-873041).
\end{acks}

\bibliographystyle{ACM-Reference-Format}
\bibliography{references}

\appendix

\end{document}